\documentclass[lettersize,journal]{IEEEtran}
\usepackage{amsmath,amsfonts}
\usepackage{algcompatible}
\usepackage{algorithm}
\usepackage{array}
\usepackage{textcomp}
\usepackage{stfloats}
\usepackage{url}
\usepackage{verbatim}
\usepackage{graphicx}
\usepackage{cite}
\usepackage{xcolor}
\usepackage{bm}
\usepackage{hhline}
\usepackage{multirow}
\usepackage{multicol}
\usepackage{siunitx}
\usepackage{caption}
\usepackage{hyperref}
\usepackage{subcaption}
\usepackage{threeparttable}
\usepackage{makecell}
\usepackage{tikz}

\newtheorem{remark}{Remark}

\definecolor{ao(english)}{rgb}{0.0, 0.5, 0.0}

\DeclareMathOperator*{\argmax}{arg\,max}
\def\nz{{n_\text{z}}}   
\def\ny{{n_\text{y}}}   
\def\ns{{M}}    
\def\Np{{N_p}}  
\def\Nmpc{{N}}  
\def\ngrid{{n_\text{G}}}
\def\ndict{{M}}
\def\xnom{{\tilde{\bm \xi}}}    

\begin{document}

\title{Active Exploration in Iterative Gaussian Process Regression for Uncertainty Modeling in Autonomous Racing}

\author{Tommaso Benciolini$^{1}$, Chen Tang$^{2}$, Marion Leibold$^{1}$, Catherine Weaver$^{3}$, Masayoshi Tomizuka$^{3}$, Wei Zhan$^{3}$
\thanks{Manuscript received Month Day, Year; revised Month Day, Year.}
\thanks{$^{1}$T. Benciolini and M. Leibold are with the Chair of Automatic Control Engineering at the Technical University of Munich, Germany (email: {\tt\small\{t.benciolini;marion.leibold\}@tum.de}). This work was developed during T. Benciolini's visit to the University of California, Berkeley.}
\thanks{$^{2}$C. Tang is with the Department of Computer Science at the  University of Texas at Austin, TX, USA (email: {\tt\small chen.tang@utexas.edu}). Work done when C. Tang was with the University of California, Berkeley.}
\thanks{$^{3}$C. Weaver, M. Tomizuka, and W. Zhan are with the Department of Mechanical Engineering at the University of California, Berkeley, CA, USA (email: {\tt\small\{catherine22;tomizuka;wzhan\}@berkeley.edu}).}
}

\maketitle

\begin{abstract}
Autonomous racing creates challenging control problems, but Model Predictive Control (MPC) has made promising steps toward solving both the minimum lap-time problem and head-to-head racing. Yet, accurate models of the system are necessary for model-based control, including models of vehicle dynamics and opponent behavior. Both dynamics model error and opponent behavior can be modeled with Gaussian Process (GP) regression. GP models can be updated iteratively from data collected using the controller, but the strength of the GP model depends on the diversity of the training data. We propose a novel active exploration mechanism for iterative GP regression that purposefully collects additional data at regions of higher uncertainty in the GP model. In the exploration, a MPC collects diverse data by balancing the racing objectives and the exploration criterion; then the GP is re-trained. The process is repeated iteratively; in later iterations, the exploration is deactivated, and only the racing objectives are optimized. Thus, the MPC can achieve better performance by leveraging the improved GP model. We validate our approach in the highly realistic racing simulation platform Gran Turismo Sport of Sony Interactive Entertainment Inc for a minimum lap time challenge, and in numerical simulation of head-to-head. Our active exploration mechanism yields a significant improvement in the GP prediction accuracy compared to previous approaches and, thus, an improved racing performance.
\end{abstract}


\section{Introduction}
\IEEEPARstart{A}{mong} the many applications of autonomous driving, autonomous racing has recently gained increased attention in research~\cite{betz2022}, also for real-world tests like Roborace and the Indy Autonomous Challenge. Two scenarios are considered: the minimum time trial and head-to-head racing against an opponent. In the former, a single race car drives around a constrained track trying to minimize the lap time. In this scenario, the main control challenges arise from pushing the vehicle to the handling limits, a task that expert humans can do well, but is challenging for control algorithms. In particular, physics-only models typically used in urban or highway environments are not well suited to represent the vehicle dynamics close to the handling limits. Within model-based controllers, Model Predictive Control (MPC) relies on a prediction model of the vehicle, and the input is determined by iteratively solving an optimal control problem over a finite horizon. Thus, the large uncertainty introduced by the modeling errors when the vehicle is driven near handling limits must be accounted for, for example adding a learning component to the physics-based model~\cite{hewing2018,rosolia2017,rosolia2020}. Furthermore, it has been shown that model-free reinforcement learning can outperform human performance~\cite{fuchs2021}. However, a main challenge with learning-based methods is to obtain data sufficiently representative, while still avoiding dangerous situations.

In the scenario with an opponent, the controlled vehicle, named Ego Vehicle (EV) in this work, must compete with another agent and perform overtaking maneuvers. In this case, the uncertainty about vehicle dynamics near the handling limits is compounded by another major challenge: the interaction with the other agent, which is a well-researched problem for autonomous urban and highway driving~\cite{song2020,liu2021}. Both enforcing collision avoidance and planning a successful overtaking maneuver require the EV to handle the uncertainty around the unknown future position of the opponent. Initial approaches considered \textit{passive} prediction models, that is, predicting the future trajectory of other agents given historical data and the current traffic configuration. Such approaches allow for a simplified planning framework, in which the future trajectories of other agents are assumed to be independent of the current decision of the EV. However, in highly interactive scenarios, such as automated racing, where the other agent is a competing opponent, the reaction of other agents to the EV decision must be considered. Knowledge of the opponent's reaction to its own future trajectory is crucial to allow overtaking maneuvers. To account for the reaction to own decisions, the opponent can be represented as a rational agent in a game-theoretic framework~\cite{wang2019}, which is, however, computationally demanding. Alternatively, the policy of the opponent to the current and past configurations and the EV own decision can be learned from past data~\cite{brudigam2021d,zhu2022}. In the latter case, it is fundamental to retrieve a training dataset sufficiently representative to allow for reliable learning of the policy.

In this work, we address both sources of uncertainty in autonomous racing, i.e., modeling errors in the dynamics and the representation of the unknown policy of the opponent accounting for the reaction to the EV's own decisions, using an iterative Gaussian Process (GP) regression algorithm, following the approach from~\cite{su2023}. GP regression is a non-parametric framework to make predictions of an unknown function given a prior and a dataset of previously collected measurements. Our main contribution is extending the iterative GP framework by adding an active exploration mechanism designed to retrieve representative data and improve learning performance. Previous works in autonomous racing did not consider the active exploration of the feature space to improve the learning performance and relied on data collected while maximizing the EV performance. Compared to other learning tools, such as artificial neural networks, a major advantage of GPs is that a measure of the model uncertainty is provided, which we exploit in the exploration mechanism to yield an enriched dataset and, ultimately, an improved prediction. In our algorithm, the dataset of the GP is updated iteratively with the measurements collected over several runs, re-training the model when the dataset is updated. During the initial iterations, the reference trajectory of the EV is designed to encourage the exploration of the regions of the feature space with a high posterior covariance of the prediction error. In doing so, the dataset is rapidly replaced with properly selected data points that refine the learning performance.

We find that enriching the dataset through the active exploration mechanism yields a significant improvement in the learning performance and, eventually, in the EV performance. We show that the GP exploration algorithm can be applied successfully both when the GP model is used for error compensation in the minimum lap time task and for the opponent modeling in head-to-head racing. For the minimum lap time, we test the algorithm in the highly realistic racing simulation platform Gran Turismo Sport of Sony Interactive Entertainment Inc~\cite{SonyGTS}, where high-fidelity dynamics models are used to simulate the vehicle, and we offer a comparison with the previous work~\cite{su2023}. For the task with the opponent, we compare our algorithm with the previous work~\cite{zhu2022} in the simulation environment therein provided.

The contributions of this work are as follows:
\begin{itemize}
    \item We propose an iterative GP regression framework with an active exploration mechanism to explore the most uncertain regions of the state space as indicated by the GP posterior covariance matrix;
    \item We implement the active exploration mechanism in the objective of the model predictive controllers for both a time trial race and a head-to-head racing challenge;
    \item We show that our method, which combines uncertainty-based exploration with training dataset selection of the most diverse data,  improves the GP prediction accuracy and the EV racing performance in comparison simulations with previous approaches from the literature.
\end{itemize}

In Section~\ref{sec:related_work}, we review the relevant related work, whereas Section~\ref{sec:vehicle_model} and~\ref{sec:GP} present preliminaries regarding the vehicle model and GP regression, respectively. Our novel iterative GP regression framework with active exploration is presented in Section~\ref{sec:active_mechanism}, detailing the procedure for the time trial and for the opponent challenge in Section~\ref{sec:method_min_lap_time} and~\ref{sec:method_opponent}, respectively. The validation simulations for both autonomous racing scenarios are presented and analyzed in Section~\ref{sec:simulations}. The conclusion with an outlook for future research is given in Section~\ref{sec:conclusion}.

\subsection{Related Work}
\label{sec:related_work}
In this section, we give a concise review of the relevant related work concerning GP regression for the compensation of vehicle dynamics and opponent modeling in autonomous racing, as well as existing approaches for active exploration presented in other fields. Further relevant works on automated racing can be found in the recent survey~\cite{betz2022}.

\subsubsection*{\textbf{Vehicle Dynamics and GP Compensation Models}}
A variety of models to describe the vehicle dynamics have been proposed in literature~\cite{rajamani2011,kong2015}. In autonomous urban and highway driving, the bicycle model~\cite{rajamani2011} referred to the Frenet-Serret coordinate system is a popular choice, representing a good trade-off between low complexity and accuracy for such applications. Nevertheless, racing vehicles operate in highly nonlinear regions of the vehicle dynamics, for which such model alone proves considerably inaccurate, and phenomena such as drifting and weight transfer must be taken into account~\cite{hao2022}. Even well-known models such as Pacejka’s Magic Formula~\cite{pacejka2005}, which models the empirical tire friction, are not well suited to represent vehicle behavior in extreme conditions. Therefore, research has focused on learning-based approaches. Although completely model-free methods have been applied successfully~\cite{wurman2022}, the lack of interpretability of deep-learning models makes it challenging to analyze the learned policy. Furthermore, neural networks do not provide a measure of the uncertainty around the prediction, which is necessary to determine the regions of the feature space that are not well represented and that must be explored.

Gaussian Process regression is a powerful tool, which also provides the posterior covariance of the prediction, and has been extensively used in autonomous racing. In~\cite{kabzan2019}, a GP is trained with data points collected during the operation. The dataset is updated online to replace previous measurements depending on a policy of temporal vicinity and information gain. In~\cite{hewing2018}, a GP is similarly used as a residual model, further taking advantage of the model uncertainty yielded by the GP to formulate chance constraints to improve safety. An approach to improve the tracking performance by means of a GP prediction model is presented in~\cite{jain2021}, in which previous measurements are used to train the compensation model. An iterative GP regression in combination with MPC is used in~\cite{hao2022} to improve the online tracking performance using data from the previous iterations. The approach is further extended in~\cite{su2023} iteratively retraining a GP used as modeling error compensation at the planning level since model inaccuracies affect also the optimal planned path. However, measurements collected during normal EV operation might not be sufficiently representative of the EV dynamics in all situations, and the GP compensation model can be further improved. This especially relates to the GP compensation for the planning algorithm, in which it is crucial to have an accurate representation of the EV dynamics in all regions of the track. However, the measurements collected during normal operation are not sufficiently diverse in general, and an exploration mechanism is necessary to improve the prediction accuracy. In~\cite{wischnewski2019}, a model-free learning method based on GP regression is used to compensate for the vehicle dynamics in autonomous racing. The bounds on the maximum accelerations in longitudinal and lateral directions are progressively updated based on the collected measurements, expanding the range of feasible acceleration values that are likely to be safe. However, the exploration of specific uncertain regions of the feature space is not actively encouraged.

\subsubsection*{\textbf{Opponent Modeling}}
The autonomous racing scenario with an opponent has been considered in~\cite{wang2019} using a game-theoretic framework, in which the policy of the EV is chosen as a Nash equilibrium, following well-established approaches for urban and highway autonomous driving~\cite{bahram2016,dreves2018,laine2021,evens2022}. However, the solution of a dynamic game is generally a computationally expensive task. Moreover, accurate knowledge of the opponent's own reward function and constraints is required to implement this approach, which is limiting in practice. Alternatively, machine learning methods have been used to directly learn the policy or the closed-loop trajectory of the opponent from data. In~\cite{brudigam2021d}, a GP is used to learn a mapping from the current EV and opponent state to the future opponent state and the posterior covariance of the GP is used to tighten safety collision avoidance constraints. The approach is interesting and relatively computationally inexpensive at run time, however implements a passive interaction approach, in which the reaction of the opponent to the current EV's own decisions is not considered. In~\cite{espinoza2022}, in the context of urban autonomous driving, a neural network is used to approximate the closed-loop behavior of other agents in a game-theoretic fashion. Instead of solving an optimization problem to predict the future trajectory of other agents, their reaction to the EV's own decision is predicted by a neural network that takes as input the future state of the EV. However, a neural network does not provide a measure of the uncertainty around the prediction. In~\cite{zhu2022}, a GP is trained in a similar fashion, conditioning on the future plan of the EV as well. However, the model is trained on a dataset of measurements collected during normal operation in several previous runs. As a result, the GP prediction of the opponent is not accurate for all possible overtaking strategies that the EV can attempt. To improve the prediction accuracy, an active exploration mechanism is needed, explicitly targeting more regions of the feature space. 

\subsubsection*{\textbf{Active Exploration}}
For learning-based approaches, the choice of the training set plays a major role in determining the performance of the learned model and in the generalization capability. In particular, in iterative approaches, in which the data used are measurements collected during the previous iterations while maximizing the performance of the EV, the dataset might not be sufficiently expressive to significantly improve the model in the whole feature space.

Approaches for active exploration in GP regression have been proposed for control of wind farms~\cite{park2016}, airborne wind energy system~\cite{bin-karim2019,siddiqui2023}, or UAV delivery control~\cite{yang2017}. In such applications, an accurate and updated estimate of the wind field is fundamental, therefore the referenced works proposed approaches to trade-off between maximizing the performance of the system and controlling it in a way to collect measurements to improve the wind field estimation.

In autonomous racing, however, compromising the performance objectives is only acceptable during the early stages of the competition, whereas eventually, the focus must be the maximization of the EV performance. Therefore, the trade-off between exploration and performance objectives must be tuned dynamically. Furthermore, the decision on the regions to be explored must take place in real-time.

\section{Preliminaries}
In this section, we detail preliminaries for our work. Section \ref{sec:vehicle_model} describes a dynamic bicycle model that will serve as a nominal vehicle model. Section \ref{sec:GP} describes Gaussian Process (GP) models, which be used in later sections for error compensation and opponent modeling.
\subsection{Vehicle Dynamics}
\label{sec:vehicle_model}
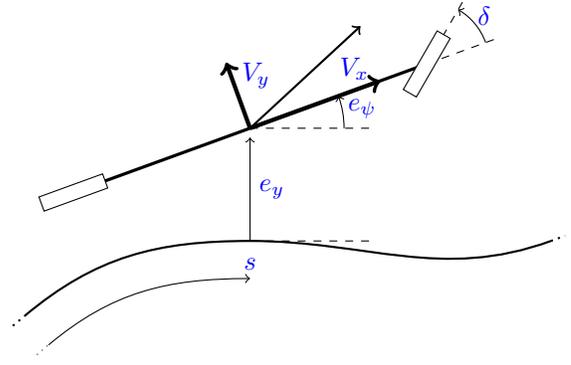
\begin{figure}
    \centering
    \begin{tikzpicture}

\def \Ax {-3}
\def \Ay {-2.5}
\def \Aang {40}
\def \Bx {0}
\def \By {-1.5}
\def \Cx {4}
\def \Cy {-1.5}
\def \Cang {200}
\def \len {5}
\def \yaw {20};
\def \steering {40};
\def \wlen {0.9};
\def \wwid {0.2};
\def \vlen {2};
\def \dotlen {0.2};
\def \corrfactor {1.2};

\node (O) at (0,0) {};

\draw[thick] (\Ax,\Ay) to  [out = \Aang, in = 180] (\Bx,\By) to [out=0,in=\Cang]  (\Cx,\Cy);
\draw[dotted, thick] ({\Ax+\dotlen*cos(180+\Aang)},{\Ay+\dotlen*sin(180+\Aang)}) to (\Ax,\Ay);
\draw[dotted, thick] (\Cx,\Cy) to ({\Cx+\dotlen*cos(180+\Cang)},{\Cy+\dotlen*sin(180+\Cang)});

\draw[->] ({\Ax+0.5*sin(\Aang)},{\Ay-0.5*cos(\Aang)}) to [out = \Aang, in = 180] ({\Bx},{\By-0.5});
\draw[dotted] ({\Ax+0.5*sin(\Aang)+\dotlen*cos(180+\Aang)},{\Ay-0.5*cos(\Aang)+\dotlen*sin(180+\Aang)}) to ({\Ax+0.5*sin(\Aang)},{\Ay-0.5*cos(\Aang)});
\node at ({\Bx},{\By-0.3}) {$\color{blue}s$};

\draw[->] (\Bx,\By) to node[right,pos=0.5] {$\color{blue}e_y$} (O);

\draw[dashed] (\Bx,\By) to ({\Bx+\len/3},\By);
\draw[dashed] (0,0) to ({\len/3},0);
\draw[->] (\len/4,0) arc (0:\yaw:\len/4);
\node at ({\len/3.3*cos(\yaw/2)},{\len/3.3*sin(\yaw/2)}) {$\color{blue}e_\psi$};

\draw[very thick] ({\len/2*cos(180+\yaw)},{\len/2*sin(180+\yaw}) to ({\len/2*cos(\yaw)},{\len/2*sin(\yaw});

\draw[rotate around={\yaw:({\len/2*cos(180+\yaw)},{\len/2*sin(180+\yaw})},fill=white] ({\len/2*cos(180+\yaw)-\wlen/2},{\len/2*sin(180+\yaw)-\wwid/2}) rectangle ++(\wlen,\wwid);

\draw[dashed] ({\len/2*cos(\yaw)},{\len/2*sin(\yaw)}) to ({\len/2*cos(\yaw)+\len/5*cos(\yaw)},{\len/2*sin(\yaw)+\len/5*sin(\yaw)});
\draw[dashed] ({\len/2*cos(\yaw)},{\len/2*sin(\yaw)}) to ({\len/2*cos(\yaw)+\len/5*cos(\steering+\yaw)},{\len/2*sin(\yaw)+\len/5*sin(\steering+\yaw)});
\draw[->] ({\len/2*cos(\yaw)+\len/6*cos(\yaw)},{\len/2*sin(\yaw)+\len/6*sin(\yaw)}) arc (\yaw:\yaw+\steering:\len/6);
\node at ({\len/2*cos(\yaw)+\len/5*cos(\yaw+\steering/2)},{\len/2*sin(\yaw)+\len/5*sin(\yaw+\steering/2)}) {$\color{blue}\delta$};
\draw[rotate around={\steering+\yaw:({\len/2*cos(\yaw)},{\len/2*sin(\yaw})},fill=white] ({\len/2*cos(\yaw)-\wlen/2},{\len/2*sin(\yaw)-\wwid/2}) rectangle ++(\wlen,\wwid);

\draw[thick,->] (0,0) to ({\vlen*cos(\yaw+atan(0.5*tan(\steering)))},{\vlen*sin(\yaw+atan(0.5*tan(\steering)))});

\draw[ultra thick,->] (0,0) to node[above,pos=0.8] {$\color{blue}V_x$} ({\vlen*cos(atan(0.5*tan(\steering)))*cos(\yaw))},{\vlen*cos(atan(0.5*tan(\steering)))*sin(\yaw)});

\draw[ultra thick,->] (0,0) to node[right,pos=0.8] {$\color{blue}V_y$} ({\corrfactor*\vlen*sin(atan(0.5*tan(\steering)))*cos(\yaw+90))},{\corrfactor*\vlen*sin(atan(0.5*tan(\steering)))*sin(\yaw+90)});

\end{tikzpicture}
    \caption{Scheme of the Dynamic Bicycle model.}
    \label{fig:bicycle_model}
\end{figure}
We model the racing vehicle using a dynamic bicycle model~\cite{kong2015} referred to the road-aligned Frenet coordinates, as represented in Figure~\ref{fig:bicycle_model}. The state of the vehicle is $\bm \xi = [V_x, V_y, \dot{\psi}, e_\psi, e_y, s]^\top$, where $V_x$ and $V_y$ are the vehicle's longitudinal and lateral velocity, respectively, in the vehicle's body frame, $\dot{\psi}$ is the yaw angular velocity, $e_\psi$ and $e_y$ are the yaw angle and lateral displacement of the center of gravity of the vehicle with respect to the reference path, and $s$ represents the traveled distance along the reference path. The relative yaw angle $e_\psi$ and lateral distance $e_y$ in Frenet coordinates are defined with respect to the closest point of the reference path. The input $\bm u=[\delta, a_x]^\top$ consists of the front steering angle and of the longitudinal acceleration resulting from the powertrain, which is applied to the rear wheel.

The dynamics is derived from force-mass and inertia-moment balance, for the first two components $V_x$ and $V_y$, and
then from the kinematics for the other state quantities. The dynamics is

\begin{equation}
    \dot{\bm \xi} = 
    \begin{bmatrix}
    \displaystyle a_x-\frac{F_{y\text{f}}\sin(\delta)+R_x+F_{x\text{w}}}{m}-g\sin(\varphi)+\dot{\psi}V_y\\[2ex]
    \displaystyle\frac{F_{y\text{f}}\cos(\delta)+F_{y\text{r}}}{m}-\dot{\psi}V_x\\[2ex]
    \displaystyle\frac{l_\text{f}F_{y\text{f}}\cos(\delta)-l_\text{r}F_{y\text{r}}}{I_{zz}}\\[2ex]
    \displaystyle\dot{\psi}-\frac{V_x\cos(e_\psi)-V_y\sin(e_\psi)}{1-\kappa(s)e_y}\kappa(s)\\[2.5ex]
    V_x\sin(e_\psi)+V_y\cos(e_\psi)\\[2ex]
    \displaystyle \frac{V_x\cos(e_\psi)-V_y\sin(e_\psi)}{1-\kappa(s)e_y}
    \end{bmatrix}
    ,
    \label{eqn:dynamic_bicycle_model}
\end{equation}
where $m$ is the mass of the vehicle, $I_{zz}$ is the moment of inertia, and $l_\text{f}$ and $l_\text{r}$ represent the distance of the center
of gravity from the front and rear axle, respectively. $R_x$ is the tire rolling resistance, and $F_{x\text{w}}$ is the wind drag force applied on the vehicle body. $F_{y\text{f}}$ and $F_{y\text{r}}$ are the lateral tire forces of the front and rear tires, which are nonlinear and vary as the tire slips along the road surface. Furthermore, gravity is acting on the vehicle with acceleration $g$ and $\varphi$ is the inclination of the road. $\kappa(s)$ is the curvature of the reference path at position $s$. A more thorough discussion of the model is reported in~\cite{hao2022}.

In this work, we employ a discretized version of the model obtained via forward Euler:
\begin{equation}
    \bm \xi^+=\bm \xi+\bm{f}(\bm \xi,\bm u)T,
    \label{eqn:nominal_discrete_time_bicycle}
\end{equation}
where $T$ is the sampling time and $\bm{f}(\bm \xi,\bm u)$ is a compact representation of~\eqref{eqn:dynamic_bicycle_model}.

However, the bicycle model in~\eqref{eqn:dynamic_bicycle_model} can be insufficient to reliably describe the vehicle motion at its handling limits, which is necessary for achieving the fastest lap time possible. For instance, the single track model does not accurately represent all components of a race car, such as weight transfer and vehicle suspension, that can be crucial to improve vehicle performance~\cite{hao2022,su2023}.

\subsection{Gaussian Processes}
\label{sec:GP}
GP regression is a machine learning method used to infer the value of an unknown function given a dataset of $\ns$ measurements $\mathcal{D}=\{\bm z_i,\bm y_i\}_{i=1}^\ns$, with $\bm z_i\in\mathbb{R}^\nz$ the input features and $\bm y_i\in\mathbb{R}^\ny$ the output features. A GP is defined as a collection of random variables, each subset of which is jointly normally distributed, and is fully specified by the prior mean and the kernel used as prior covariance~\cite{rasmussen2005}. It is assumed that the underlying unknown function $\bm g(\cdot)$ relates the input and the output features as follows
\begin{equation}\label{eq:gpoutput}
    \bm y_i = \bm g(\bm z_i)+\bm w_i,
\end{equation}
where $\bm w_i\in\mathbb{R}^\ny,\bm w_i\sim\mathcal{N}(\bm 0, \bm \Sigma^{\bm w})$ is i.i.d. Gaussian noise with diagonal covariance matrix $\bm \Sigma^{\bm w}=\text{diag}(\sigma_1^2,\dots,\sigma_\ny^2)$. The unknown function is specified through its mean, which we assume zero without loss of generality, and a kernel function $k^a(\bm z,\bm z')$, where $\bm z,\bm z'\in\mathbb{R}^\nz$ are two input GP input feature vectors. The scalar function $k^a(\bm z,\bm z')$ is chosen to encode the prior assumptions and the function properties. To approximate the modeling error in the dynamics, we use the squared exponential kernel~\cite{rasmussen2005}
\begin{equation}
    k^a(\bm z,\bm z') = \sigma^2_{k^a}\exp\left(-\frac{1}{2}(\bm z-\bm z')^\top\bm L^{-2}_{k^a}(\bm z-\bm z')\right),
\end{equation}
with parameter $L_{k^a}$ defining the characteristic length-scale and $\sigma^2_{k^a}$ the squared signal variance, whereas to infer the future trajectory of the opponent, we employ the Matérn kernel with parameter $\nu=1.5$~\cite{rasmussen2005}
\begin{equation}
    k^a(\bm z,\bm z')=\left(1+\frac{\sqrt{3}\lVert\bm z-\bm z'\rVert^2}{l_{k^a}}\right)\exp\left(-\frac{\sqrt{3}\lVert\bm z-\bm z'\rVert^2}{l_{k^a}}\right),
\end{equation}
where $l_{k^a}$ is a length scale parameter. Both kernels are widely used and have been chosen consistently with~\cite{su2023} and~\cite{zhu2022}, respectively, to facilitate the comparison. The parameters of the kernels are optimized by maximizing the marginal likelihood of the observations~\cite{rasmussen2005}.

The posterior mean and covariance of $d$-th entry $g_d(\bm z)$ of the underlying unknown function $\bm g(\bm z)\sim\mathcal{N}(\bm \mu(\bm z),\bm\Sigma(\bm z))$ at the arbitrary point $\bm z^\ast$ conditioned on the training set $\mathcal{D}$ are obtained as
\begin{subequations}\label{eq:trainGP}
\begin{align}
    \mu_d(\bm z^\ast) &=(\bm k^a)^\top\bm K^{-1}\bm\gamma_d\\
    \sigma_d(\bm z^\ast) &=k^{a\ast}-(\bm k^a)^\top\bm K^{-1}\bm k^a,\label{eqn:gp_prediction_covariance}
\end{align}
\end{subequations}
where $\bm k^a=[k^a(\bm z_1,\bm z^\ast),\dots,k^a(\bm z_\ns,\bm z^\ast)]^\top$, the entries of matrix $\bm K$ are $K_{ij}=k^a(\bm z_i,\bm z_j)$, $k^{a\ast}=k^a(\bm z^\ast,\bm z^\ast)$, and $\bm\gamma_d=[y_{1,d},\dots,y_{\ns,d}]^\top$ contains the training outputs corresponding to the $d$-th entry.

\section{Method}
\label{sec:method}

In Section \ref{sec:active_mechanism}, we first introduce a general active exploration framework that can iteratively train Gaussian Process (GP) models. The exploration mechanism is augmented into the objective function of a model predictive controller (MPC), and it strategically explores locations of the state space where the GP model has the highest uncertainty. While the active exploration framework is applicable for any MPC that employs a GP model for some aspect of constraint modeling, we specifically apply the framework to two autonomous racing MPCs. The time trial racing challenge is detailed in Section~\ref{sec:method_min_lap_time}, where a GP model is used to compensate for EV modeling error in an offline optimal trajectory planner and online trajectory-tracking MPC. Then, in Section~\ref{sec:method_opponent}, we apply our framework to head-to-head racing with an opponent, where the GP model is used to predict the behavior of the opponent and the EV employs an online MPC for trajectory planning and control.

\subsection{Active Exploration and Data Selection for Iterative GP Regression}\label{sec:active_mechanism}

Our method leverages the uncertainty in the GP prediction to actively collect data in regions of high uncertainty, with the goal of improving the prediction accuracy of the GP. We assume that the control algorithm is an MPC and that some parts of the MPC's model or constraints are modeled with a GP. 
In the first iterations of our algorithm, active exploration takes place; namely, the EV control is determined as a trade-off between the MPC performance objectives and the exploration objectives.  In doing so, the enriched data can be used iteratively to re-train the GP; thus, the GP will be more accurate, particularly in regions of high uncertainty in previous iterations. Since modeling accuracy can heavily influence the performance of model-based control, or goal is to improve the overall performance of the MPC after exploration is complete, and the MPC may fully exploit the more accurate GP model. In the following parts, we discuss the components and main aspects of our proposal.

\subsubsection*{\textbf{GP Model}}
In general, the GP model predicts an unknown value, $\bm y$, as a function of the current state according to an unknown function~\eqref{eq:gpoutput}. For example, $\bm y$ may represent a compensation term for the modeling error dynamics or a prediction of the opponent's future trajectory. 

To reduce the dimensionality of the GP, rather than conditioning the GP to be a function of the full state, we can condition on an input feature vector, $\bm z$, that is a deterministic, known function of the state, i.e. $\bm z =\bm \phi_z(\bm \xi)$, where $\bm \phi_z$ is a feature extractor and $\bm \xi$ is the state of the system. The GP models the output function $\bm g^{\bm \xi}(\bm z)$ as the normal distribution
\begin{equation}\label{eq:gpmodel}
    \bm g^{\bm \xi}(\bm z)\sim\mathcal{N}\left(\bm\mu^{\bm \xi}(\bm z),\bm\Sigma^{\bm \xi}(\bm z)\right).
\end{equation}
The mean and standard deviation of the distribution, $\bm\mu^{\bm \xi}(\bm z)$ and $\bm\Sigma^{\bm \xi}(\bm z)$, are obtained from~\eqref{eq:trainGP} to predict the output $\bm y_i$ from the input feature vector $\bm z_i$. Once the GP model is trained, the controller can employ $\bm y = \bm\mu^{\bm \xi}$ to model the unknown function \eqref{eq:gpoutput} during online control.
 
\subsubsection*{\textbf{Diverse Data Selection}}
During the repeated trials, a large number of data points is collected. Using all such data to train the GP model is impractical and unnecessary, as a smaller dataset of appropriately selected data points suffices to represent the input-output relation. However, creating a smaller dataset by randomly sampling from the collected data points does not guarantee that the diverse data points collected during the exploration phase are appropriately exploited. For this reason, at the end of each iteration, we train the GP using a smaller dataset of points obtained with the data selection approach described in~\cite{nguyen-tuong2011}, outlined in the following.

The goal is to select a (small) collection of points $\mathcal{D}=\{(\bm z_i, \bm y_i)\}_{i=1}^{\ndict}$ to represent the feature space and allow GP predictions as accurate as possible. The policy to add a data point $(\bm z_i, \bm y_i)$ to the dataset or replace existing ones leverages on a similarity measure between the new data point and the present collection, namely, the posterior prediction covariance~\eqref{eqn:gp_prediction_covariance} at $\bm z_i$ given all other data points in the dataset $\mathcal{D}$. A large value of the posterior covariance signals that the location of the input feature $\bm z_i$ is poorly represented by other data points. The policy to update the dataset works as follows:
\begin{itemize}
    \item A new datum point is added if its posterior covariance given the current dataset is larger than the median of the posterior covariance of all data points currently in the dataset for at least one of the output features since the GP models considered in this work have multi-dimensional output features;
    \item If the dataset is full, the new data point replaces the data point in the dataset with the smallest posterior covariance.
\end{itemize}

If the data points yielding the lowest posterior covariance for different output dimensions are different, we consider the dimension in which the ratio between the new posterior covariance of the new point and the minimum posterior covariance of points in the dataset is the largest. Moreover, we use the outlier rejection mechanism described in~\cite[Section V-B]{kabzan2019}.

In contrast to~\cite{kabzan2019}, we do not consider a decay factor to encourage the removal of older data points first, since the goal is to maintain the most diverse data points. Older data points that have been collected during the exploration in previous iterations are, in general, more significant than recent points collected during the last iterations, in which the focus is on the maximization of the performance.
\subsubsection*{\textbf{MPC}}
At each iteration, an MPC is used to control the system and perform a trial of the task, collecting additional experimental data. The MPC optimization problem includes a value $\bm y$, which is some unknown function of the state $\bm \xi$ and modeled by a GP using $\bm y = \bm\mu^{\bm \xi}$. Previous works have shown that iteratively collecting data with the MPC and retraining the GP model~\eqref{eq:gpmodel} can improve the performance of the GP prediction and MPC. We further propose a mechanism that will employ the knowledge of the fact that $\bm y$ is predicted by a GP to purposefully explore regions of the state space where the prediction of $\bm y$ is more uncertain. Inspired by our approach for diverse data selection that follows \cite{nguyen-tuong2011}, we use large posterior covariance of the GP to indicate that those regions of the state space need further exploration.

Specifically, at each sampling time, the following optimal control problem is solved to determine the optimal sequence of input $\{\bm u_k\}_{k=0}^{\Nmpc-1}$, where $\Nmpc$ is the MPC prediction horizon:
\begin{subequations}\label{eq:mpc}
\begin{alignat}{2}
    \min_{\{\bm u_k\}_{k=0}^{\Nmpc-1}}& J\bigg(\{\bm \xi_k\}_{k=0}^{\Nmpc}, \{\bm u_k\}_{k=0}^{\Nmpc-1}, &&\alpha,\bm\Sigma^{\bm \xi}(\cdot)\bigg)\label{eqn:cost_function_extended}\\
    \text{s.t. }\bm \xi_{k+1}&=\bm{f}(\bm \xi_k,\bm u_k)+\bm\mu^{\bm \xi},\ &&\forall k=0,\dots,\Nmpc-1\label{eqn:dynamics_extended}\\
 \bm \xi_k&\in\Xi_k,\ \ &&\forall k=1,\dots,\Nmpc\label{eqn:state_constr_extended}\\
    \bm u_k&\in\mathcal{U}_k,\ &&\forall k=0,\dots,\Nmpc-1.\label{eqn:input_constr_extended}
\end{alignat}
\end{subequations}
The MPC cost function, $J$, in ~\eqref{eqn:cost_function_extended} is a function of the predicted states and control actions \emph{and} the posterior covariance of the GP model,  $\bm\Sigma^{\bm \xi}(\cdot)$. The cost function will trade-off between the MPC's original performance objectives and exploration objectives using the weighting parameter, $\alpha$ according to the \textbf{\textit{Active Exploration}} algorithm described in the next paragraph. Additionally, Eq.~\eqref{eqn:dynamics_extended} constrains the states and controls to follow the dynamic model of the system, where the unknown components of the dynamics are compensated for by the term $\bm \mu^{\bm \xi}$, which is the predicted output of the GP: $\bm \mu^{\bm \xi}=\bm y$. The constraint sets for the state, $\Xi$ in~\eqref{eqn:state_constr_extended}, and control, $\mathcal{U}$ in~\eqref{eqn:input_constr_extended}, restrict the state and control based on dynamics and any additional constraints such as collision avoidance. To account for the uncertainty about the prediction $\bm y$, state constraints~\eqref{eqn:state_constr_extended} are tightened for the components of the system state that are affected by uncertainty, as we discuss in Section~\ref{sec:method_min_lap_time} and~\ref{sec:method_opponent} for each racing scenario.

\subsubsection*{\textbf{Active Exploration}}

The goal of the active exploration mechanism is to solve  \eqref{eq:mpc} such that it encourages the exploration of the feature space and collects new measurements that enrich the dataset  $\mathcal{D}$.  We assume that one of the objectives of the cost function $J$ is to minimize the state's distance to a reference state $\bm \xi_k^\text{ref}$, which is common practice for optimal control problems in racing \cite{hao2022}. This allows us to propose a mechanism to set the reference $\bm \xi_k^\text{ref}$ so that the MPC targets states in $\bm z^{\text{ref}}$, where the uncertainty in the prediction $\bm y$ is large. 

We define a target value,  $\bm z^{\text{ref}}$, that includes states used for GP prediction that the MPC should explore to improve the prediction accuracy of the GP. The target value $\bm z^{\text{ref}}$ balances criteria based on the true current state and the exploration objective. 
It is chosen from a list of $\ngrid$ candidate values, $\{\bm z^{(i)}\}_i^{\ngrid}$, that are created from typical values for each feature that are combined to create the multi-dimensional feature set in a combinatorial fashion. Each state's typical values are obtained from previous data.

Algorithm~\ref{alg:feature_selection} describes our method for selecting $\bm z^{\text{ref}}$ from the candidate target values and is inspired by the use of mutual information to select features worth exploring. Algorithm~\ref{alg:feature_selection} takes as input the state reference based on the performance objective $\bm \xi_k^\text{ref}$, for example, the trajectory computed by the planner in the time trial application. At the end of the Algorithm, the state reference based on the performance objective $\bm \xi_k^\text{ref}$ is updated based on the selection of the candidate target values $\bm z^{\text{ref}}$. The method is based on previous work~\cite{yang2017}, where instead, the knowledge of mutual information is used. We use the weighted posterior covariance that is consistent with the data selection algorithm and to allow to specify the priority within the states during the exploration.

First, $\tilde{\bm z}$ is computed, that is, the features of the state that would be visited if the MPC follows only the performance-based objectives.
Then, the candidate features are ordered based on two competing criteria: 1.) their proximity $d_i$ to feature corresponding to the performance-objectives $\tilde{\bm z}$ (lines 3-5) and 2.) the posterior covariance $v_i$ calculated using the current iteration's GP model and weighted by the matrix $S$ (lines 6-8). $D_i$ and $V_i$ on lines 5 and 8 represent the position of the $i$-th candidate input feature vector in the list ordered by distance and in the list ordered by variance, respectively.  To trade-off between these objectives, we select the target feature by maximizing their convex combination with weight $\alpha\in[0,1]$ (line 9).
Once the target value is selected, for each state in $\bm z^{\text{ref}}$, we replace that state in  $\bm \xi_k^\text{ref}$ with the target state in $\bm z^{\text{ref}}$.
Specifically, if the GP input features are extracted from the states via $\bm z=\bm \phi_z(\bm \xi)$, we define $\bm \xi=\bm \phi_z^{-1}(\bm z)$ to indicate that $\bm \xi$ is equal to $\bm z$ for all of the states that are available in $\bm z$. Thus, once the target value $\bm z^{\text{ref}}$ is determined, we use it to update the reference state based on the performance objectives via $\bm \xi_k^\text{ref}\leftarrow\bm \phi_z^{-1}(\bm z^{\text{ref}})$. Then, the MPC computes the control using the updated reference $\bm \xi_k^\text{ref}$.

The covariance of GP has been used previously as a mechanism for data selection~\cite{nguyen-tuong2011}, as well as for autonomous racing~\cite{kabzan2019}. Thus, by selecting $\bm z^{\text{ref}}$ to balance the distance and covariance criterion, the MPC balances exploration while remaining near the original MPC reference. Increasing $\alpha$ places more weight on the exploration of the feature space, and when $\alpha=0$, the MPC defaults to use its standard reference.
The target value $\bm z^{\text{ref}}$ should be relatively near $\tilde{\bm z}$ (small $d_i$) to prevent significant deterioration in the performance of the controlled system, and to prevent possibly dangerous behaviors and loss of stability during the exploration. Conversely, $\bm z^{\text{ref}}$ should correspond to values with large posterior covariance, $v_i$, to explore uncertain regions of the state space.
We opt to weight the posterior covariance $\bm\Sigma^{\bm \xi}(\bm z^{(i)})$ by positive semi-definite matrix $\bm S\geq0$. Through the choice of $\bm S$, higher priority can be given to features that improve the data diversity for states that are most relevant to the application scenario. It is also worth observing that the posterior covariance for each candidate feature in the list can be computed immediately after the training of the GP and stored prior to MPC run time, significantly reducing the computational demand of Algorithm~\ref{alg:feature_selection} at run time.

In the first iterations, we set $\alpha$ to be large, so that the focus is placed on the exploration of the feature space and the collection of data points that enrich the dataset. In later iterations, $\alpha$ is decreased to zero; thus, the focus is entirely on the performance objectives of the MPC, taking advantage of the accurate GP prediction model obtained from training with the dataset from the exploration.

\begin{algorithm}[t]
\caption{Active exploration via state selection for the MPC reference}\label{alg:feature_selection}
\begin{algorithmic}[1]
\STATE\small\textbf{Input}: $\bm \xi_k^\text{ref}$, $\{\bm z^{(i)}\}_i^{\ngrid}$, $\bm\Sigma^{\bm \xi}(\cdot)$ $\bm S$, $\alpha$
\STATE $\tilde{\bm z}\gets\bm \phi_z(\bm \xi_k^\text{ref})$  \textcolor{gray}{\# GP feature following performance objectives}
\STATEx \textcolor{gray}{Sort candidate features by distance to the current state's feature:}
\STATE $d_i\gets\lVert\bm z^{(i)}-\bm\tilde{\bm z}\rVert\ \forall i=1,\dots,\ngrid$ 
\STATE $\bm d\gets \text{sort}(d_1,\dots,d_{\ngrid})$ 
\STATE $D_i\gets i: \bm d[i] = d_i\ \forall i=1,\dots,\ngrid$ 
\STATEx \textcolor{gray}{Sort candidate features by their weighted posterior covariance:}
\STATE $v_i\gets\lVert\bm\Sigma^{\bm \xi}(\bm z^{(i)})\rVert^2_{\bm S}\ \forall i=1,\dots,\ngrid$ 
\STATE $\bm v\gets \text{sort}(v_1,\dots,v_{\ngrid})$ 
\STATE $V_i\gets i: \bm v[i] = v_i\ \forall i=1,\dots,\ngrid$ 
\STATEx \textcolor{gray}{Select reference features using trade-off parameter $\alpha$:}
\STATE $\bm z^{\text{ref}} = \bm z^{(i^*)}$, where $i^\ast\gets \argmax_{i}\left(\alpha V_i+(1-\alpha)D_i\right)$ 
\STATEx \textcolor{gray}{Update the reference with the states in $\bm z^{\text{ref}}$:}
\STATE $\bm \xi_k^\text{ref}\leftarrow\bm  \phi_z^{-1}(\bm z^{\text{ref}})$
\STATE\small\textbf{Output}: \textit{Updated MPC reference, }$\bm \xi_k^\text{ref}$
\end{algorithmic}
\end{algorithm}

\subsection{Minimum Lap Time Application}
\label{sec:method_min_lap_time}
In the minimum lap time task, our iterative exploration-based controller is used to compensate for modeling errors in the dynamics when the vehicle approaches handling limits. To minimize the lap time, first, a time-optimal trajectory for the EV is planned, then, the vehicle is driven around the track by the MPC. Relying on the nominal model of the vehicle dynamics does not suffice to minimize the lap time, therefore we use a GP compensation model to improve the prediction of the EV state. It is important to account for the influence of unmodeled effects of the dynamics also on the optimal path, therefore we use the GP compensation both in the planning and in the MPC tracking phase, adopting the double GP compensation scheme that was presented in~\cite{su2023}. At the end of the trial, the measurements collected are used to retrain both GP models, and the trials are repeated iteratively. In the following, we outline the details of our active exploration mechanism, which is implemented in the online tracking phase. The details of the derivation of the time-optimal reference are given in the Appendix for completeness.

\subsubsection*{\textbf{GP Model}}
In the case of the time trial, the controlled system state $\bm \xi$ is the state of the EV.
We employ a GP to compensate for the unmodeled dynamics of the vehicle \cite{su2023}. Here the predicted value is $y^{\text{MPC}}_k=\bm \xi_{k+1}-\bm \xi_{k+1}^{\text{pred}}$, where $\bm \xi_{k+1}$ is the next state and $\bm \xi_{k+1}^{\text{pred}}=\bm \xi_k+\bm f(\bm \xi_k, \bm u_k)T$ is the next state predicted by the nominal vehicle model from the current state~\eqref{eqn:nominal_discrete_time_bicycle}. Thus, during control,  $\bm y^{\text{MPC}}_k$ can be added to $\bm \xi_{k+1}^{\text{pred}}$ to compensate for modeling errors. 
Specifically, if the GP compensation term added to the nominal system dynamics is defined as
\begin{equation}
    \bm g^\text{MPC}(\bm z_\text{MPC})\sim\mathcal{N}\left(\bm\mu^\text{MPC}(\bm z_\text{MPC}),\bm\Sigma^\text{MPC}(\bm z_\text{MPC})\right),
\end{equation}
then the system dynamics can be modeled as 
\begin{equation}
    \bm \xi_{k+1}=\bm A_k\bm\xi_k+\bm B_k\bm u_k+\bm d_k+\bm\mu^\text{MPC}(\bm z_\text{MPC}), 
\end{equation}
where $\mu^\text{MPC}$ models the error of the linearized dynamic bicycle model~\eqref{eqn:nominal_discrete_time_bicycle} with respect to the real dynamics at the GP input feature $\bm z_\text{MPC}$. The GP compensates the states having the greatest impact on the prediction error, $V_y$ and $\dot{\psi}$~\cite{su2023}, i.e., $\bm\mu^\text{MPC}(\bm z_\text{MPC})=[0,\mu^\text{MPC}_{V_y}(\bm z_\text{MPC}),\mu^\text{MPC}_{\dot{\psi}}(\bm z_\text{MPC}),0,0,0]^\top$. In this application, the GP input feature vector is $\bm z_\text{MPC}=\bm \phi_{z}^{\text{MPC}}(\bm \xi, \bm u)=[V_y,\dot{\psi},\delta]^\top$, that is, mapping $\bm \phi_{z}^{\text{MPC}}$ is computed not only from the reference state, but also from the reference input. Furthermore, defining $\bm \phi_{z}^{\text{MPC}}$ with respect to a nominal predicted trajectory rather than on the actual predicted state $\bm\xi_k$ and predicted input $\bm u_k$ allows real-time computation of the MPC~\cite{hao2022,su2023}. Precisely, $\bm z_\text{MPC}$ is computed from the nominal state $\xnom_k$  and $\tilde{\bm u}_k$ of the linearized dynamics of the previous MPC iteration. We record the values of $(\bm z^{\text{MPC}}_{k}, \bm y^{\text{MPC}}_k)$ during real-time control to fill the training dataset $\mathcal{D}$, since learning from offline data would not be possible.

Measurements that may be collected while tracking the optimal path are not sufficiently diverse to train the GP since the states will be concentrated in a small state space encountered while minimizing deviations from the reference path itself. This limits the learning performance and, consequently, the improvement yielded by the GP compensation both in the planning and MPC tracking. Therefore, we adopt our active exploration scheme aimed at enriching the dataset of measurements.

\subsubsection*{\textbf{MPC}}
The optimal control problem of the tracking MPC is
\begin{subequations}
\begin{alignat}{2}
    \min_{\{\bm u_k\}_{k=0}^{\Nmpc-1}}& \sum_{k=0}^{\Nmpc-1}\lVert\bm\xi_k-\bm\xi_k^\text{ref}\rVert_{\bm Q}+&&r_\delta\Delta\delta_k^2
    \label{eqn:cost_function_mpc}\\
    \text{s.t. }\bm \xi_{k+1}&=\bm A_k\bm\xi_k+\bm B_k\bm u_k+\bm d_k&&+\bm\mu^\text{MPC}(\bm \phi_z^\text{MPC}(\xnom_k, \tilde{\bm u}_k)),\nonumber\\
    &\ &&\forall k=0,\dots,\Nmpc-1\label{eqn:dynamics_mpc}\\
    w_{\text{r},k}+\gamma_k&\leq e_{y,k}\leq w_{\text{l},k}-\gamma_k,\ \ &&\forall k=1,\dots,\Nmpc\label{eqn:max_ey_mpc}\\
    \bm u_{\text{min},k}&\leq \bm u_k\leq \bm u_{\text{max},k},\ &&\forall k=0,\dots,\Nmpc-1,\label{eqn:max_input_mpc}
\end{alignat}
\label{prb:mpc_problem}%
\end{subequations}
where $\Nmpc$ is the MPC prediction horizon. The cost function~\eqref{eqn:cost_function_mpc} is designed to penalize rapid changes in the steering angle according to weight $r_\delta>0$, with $\Delta\delta_k = \delta_k-\delta_{k-1}$ and $\delta_{-1}$ is set equal to the last applied steering angle $\delta_{t-1}$ at time $t-1$. $\bm Q\geq0$ is the weight to penalize deviations of the state $\bm\xi_k$ from the reference $\bm\xi_k^\text{ref}$, which plays an important role in encouraging the exploration of the feature space depending on the value of $\alpha$. Constraint~\eqref{eqn:dynamics_mpc} relies on a linearized version of the bicycle model dynamics, computed with respect to a nominal trajectory $\xnom,\tilde{\bm u}$ \cite{hao2022}. Since the GP model is also not embedded in the optimization~\cite{hao2022}, minimization problem~\eqref{prb:mpc_problem} is a quadratic problem that can be solved in real time.

\subsubsection*{\textbf{Constraint Tightening}}
Constraints~\eqref{eqn:max_ey_mpc} and~\eqref{eqn:max_input_mpc} ensure that the lateral position of the vehicle and the input stay within the track bounds and the actuation bounds, respectively. Since the prediction of the lateral error in~\eqref{eqn:dynamics_mpc} is influenced by the GP compensation $\bm g^\text{MPC}(\bm z_\text{MPC})$, in contrast to~\cite{su2023} we tighten the constraints to address the uncertainty in the prediction. Taking the uncertainty into account in the constraints is crucially important to reduce the probability of dangerous movements of the EV during the exploration phase. The modeling error $\epsilon_{y,k}$ for $e_y$ at prediction step $k$ is an affine transformation of Gaussian variables, therefore, is also Gaussian distributed. Thus, the support of the uncertainty $\epsilon_{y,k}$ is unbounded, and a robust tightening, guaranteeing constraint satisfaction for all realizations of the uncertainty $\epsilon_{y,k}$, is not possible. Hence, we implement a stochastic tightening requiring
\begin{equation}
    \Pr\left[e_{y,k}+\epsilon_{y,k}\leq w_{\text{l},k}\right]\geq\beta,
    \label{eqn:ey_constraint_tightening}
\end{equation}
where $0\leq\beta\leq1$ is the risk parameter. Constraint~\eqref{eqn:ey_constraint_tightening} yields a deterministic formulation for the tightening parameter $\gamma_k$ in~\eqref{eqn:max_ey_mpc}. The covariance matrix $\bm\Sigma_k^{\bm\xi}$ of the predicted state $\bm\xi_k$ at step $k=1,\dots,\Nmpc$ is obtained recursively from the dynamics~\eqref{eqn:dynamics_mpc} and from the covariance $\bm\Sigma^\text{MPC}(\bm z_\text{MPC})$ of the GP compensation $\bm g^\text{MPC}(\bm z_\text{MPC})$ as
\begin{equation}
    \bm\Sigma_{k+1}^{\bm\xi}=\bm A_k\bm\Sigma_k^{\bm\xi}\bm A_k^\top+\bm\Sigma^\text{MPC}(\bm z_\text{MPC}),
    \label{eqn:covariance_predicted_mpc_state}
\end{equation}
where for all prediction steps $k$ the state $\bm\xi_k$ and the GP compensation $\bm g^\text{MPC}(\bm z_\text{MPC})$ are uncorrelated because the compensation is computed from a nominal trajectory, $\xnom_k,\tilde{\bm u}_k$. From $\bm\Sigma_k^{\bm\xi}$, the covariance $\sigma_{e_y,k}^2$ of the prediction error $\epsilon_{y,k}$ at prediction step $k$ is obtained and the tightening parameter $\gamma_k$ is computed as in~\cite{carvalho2014a}
\begin{equation}
    \gamma_k=\sqrt{2}\sigma_{e_y,k}\text{erf}(2\beta-1).
\end{equation}
Because of symmetry, the same tightening parameter is applied to the lower bound in~\eqref{eqn:max_ey_mpc}.

\subsubsection*{\textbf{Active Exploration}}
The reference trajectory $\bm\xi_k^\text{ref}$ in~\eqref{eqn:cost_function_mpc} consists of the (time-varying) reference of the state vector. To encourage the exploration of the feature space, $\bm\xi_k^\text{ref}$ is selected online using the procedure in Algorithm~\ref{alg:feature_selection}. The optimal state $\bm\xi^\text{plan}_k$ and input $\bm u^\text{plan}_k$ for each predicted time step of the optimal planned path is given as input to Algorithm~\ref{alg:feature_selection}, so that the target feature $\bm z^{\text{ref}}$ is selected as a trade-off between tracking the optimal planned path and visiting the most informative features, that is, those with the largest posterior covariance given the GP model used in the current iteration. As weighing matrix $\bm S$ of the feature covariance in Algorithm~\ref{alg:feature_selection}, we use the same weights as in the MPC cost function~\eqref{eqn:cost_function_mpc}, with a view at giving priority to features where the posterior covariance is larger for those components that are more relevant for the MPC tracking.

\begin{remark}
    The exploration mechanism is implemented considering the GP model used in the MPC optimal control problem, but it is important to improve the accuracy of the GP model used in the planning phase, otherwise, the reference path could not be time-optimal for the actual vehicle dynamics. Nevertheless, because of the relation between the features of the two GPs involved in the planning and tracking, respectively, the exploration procedure yields a diverse dataset for the GP model used in the planning phase. In fact, the two GP models share the same input features, whereas the output features of the planning GP are the derivative of the output features of the tracking GP. By frequently changing the target input feature, we collect diversified measurements for the planning GP compensation.
\end{remark}

At the end of each iteration, the data points collected are analyzed and the most diverse are selected to update the dataset and maintain the most significant data points collected during the exploration. By contrast, in~\cite{su2023}, the dataset to retrain the GP models is obtained considering all collected measurements in all previous iterations and sampling randomly within them. In practice, two datasets of data points are extracted and updated in parallel, one for the planning GP model $\bm g^\text{plan}$ and one for the tracking GP model $\bm g^\text{MPC}$, each considering the respective prediction errors.

\subsection{Head-to-Head Racing Application}
\label{sec:method_opponent}
In the head-to-head race, the EV needs to predict the opponent's future trajectory in order to plan overtaking maneuvers. However, the opponent's future trajectory depends on the opponent's reaction to the EV's own decision, and it would be unrealistic to assume that the opponent's policy is known, and this fact represents a source of uncertainty. Following the approach from~\cite{zhu2022}, we model the policy and dynamics of the opponent as a GP model, which is included in the EV controller. 
Thus, in the GP model, MPC formulation, and active exploration mechanism, we define an \textit{extended} state of the system $\bm \xi^\text{E}=[\bm \xi, \bm \xi^\text{O}]^\top$, which contains both the state of the ego vehicle $\bm \xi$ and of the opponent vehicle $\bm \xi^\text{O}$. 

In the following, we present the details of the active exploration mechanism, aimed to retrieve informative data about the reaction of the opponent to several overtaking attempts of the EV. Further, we discuss specific limitations that pertain to the active exploration in this racing challenge due to the fact that the EV does not have full control over the feature space, and we outline how the active exploration takes place in the setting of a single competition with the opponent.

\subsubsection*{\textbf{GP Model}}

The combined effect of the dynamics of the opponent and its policy is modeled as a GP, namely, we consider
\begin{equation}\label{eq:oppgp}
    y^\text{O}_k = \bm\xi_{k+1}^\text{O}=\bm \xi_k^\text{O}+\bm{f}(\bm \xi_k^\text{O},\bm \pi_k(\bm \xi_k^\text{O},\bm \xi_k))T,
\end{equation}
where $\bm f$ represents the real dynamics of the opponent and $\bm \pi$ represents the one-step opponent policy, which depends on the current opponent state $\bm \xi_k^\text{O}$ and on the current EV state $\bm \xi_k$, in order to incorporate the opponent reaction to the EV decisions in the prediction steps \cite{zhu2022}. We model the one-step closed-loop dynamics 
of the opponent in \eqref{eq:oppgp} with the GP
\begin{equation}
    \bm g^\text{O}(\bm z_\text{O})\sim\mathcal{N}\left(\bm\mu^\text{O}(\bm z_\text{O}),\bm\Sigma^\text{O}(\bm z_\text{O})\right).
\end{equation}
The opponent's state is
\begin{equation}
    \bm\xi^\text{O}=[s^\text{O},e_y^\text{O},e_\psi^\text{O},V_x^\text{O}]^\top.
\end{equation}
Here $s^\text{O}$ and $e_y^\text{O}$ are the longitudinal and lateral position of the opponent on the track, $e_\psi^\text{O}$ is the yaw angle with respect to the reference of the track, and $V_x^\text{O}$ is the longitudinal velocity. The GP  input features are
\begin{equation}
    \bm z_\text{O} = \bm\phi^\text{o}_z(\bm \xi^\text{E})= [s^\text{O}-s,e_y^\text{O}-e_y,e_\psi,V_x,e_\psi^\text{O},V_x^\text{O},\bar{\bm\kappa}]^\top,
    \label{eqn:opponent_input_features}
\end{equation}
that are the longitudinal and lateral distance between the EV and the opponent, the yaw angle and longitudinal velocity of both vehicles and vector $\hat{\bm\kappa}$ containing the track curvature at three look-ahead points. The GP input features~\eqref{eqn:opponent_input_features} consist only of the relative configuration of the EV and of the opponents and of their position in the curvilinear Frenet coordinates, rather than in absolute coordinates, with a view to boosting the generalization capability of the GP prediction. Furthermore, the prediction of the opponent's trajectory is obtained by averaging over many samples from the GP model as in~\cite[Algorithm 1]{zhu2022}.

\subsubsection*{\textbf{MPC}}
The EV trajectory is computed iteratively by an MPC, which solves the following optimal control problem at each sampling time:
\begin{subequations}
\begin{alignat}{2}
    \min_{\{\bm u_k\}_{k=0}^{\Nmpc-1}}& \alpha J_\text{expl}\bigg(\{\bm\xi_k\}_{k=1}^{\Nmpc},\ \{\bm\xi^\text{ref}_k&&\}_{k=1}^{\Nmpc}\bigg)+(1-\alpha)\bigg(\sum_{k=0}^{\Nmpc-1}q_ce_{y,k}^2\nonumber\\
    &+\bm u_k^\top\bm R\bm u_k+\Delta\bm u_k^\top\bm R_d&&\Delta\bm u_k-q_ss_N^2\bigg)\label{eqn:cost_function_opponent}\\
    \text{s.t. }\bm \xi_{k+1}&=\bm \xi_k+\bm{f}(\bm \xi_k,\bm u_k)T\ &&\forall k=0,\dots,\Nmpc-1\label{eqn:dynamics_opponent}\\
    s_0 &= s(\bm \xi_t)\label{eqn:opponent_s_init}\\
    s_{k+1} &= s_k+V_{x,k}T\ \ &&\forall k=0,\dots,\Nmpc-1\label{eqn:opponent_s_dynamics}\\
    w_{\text{r},k}&\leq e_{y,k}\leq w_{\text{l},k}\ &&\forall k=1,\dots,\Nmpc\label{eqn:max_ey_opponent}\\
    \bm u_{\text{min},k}&\leq \bm u_k\leq \bm u_{\text{max},k}\ &&\forall k=0,\dots,\Nmpc-1\label{eqn:max_input_opponent}\\
    \bm 0&\geq \bm h(\bm\xi_k,\bm\xi^\text{O}_k)\ &&\forall k=1,\dots,\Nmpc.\label{eqn:opponent_collision_avoidance}
\end{alignat}
\label{prb:opponent_problem}%
\end{subequations}
The cost function~\eqref{eqn:cost_function_opponent} consists of the convex combination of two terms, weighted by parameter $\alpha$: in the first term, the exploration objectives are considered, that will be discussed later and which is the only difference with respect to the optimal control problem in the baseline work~\cite{zhu2022}; in the second term, racing objectives are considered, namely penalties for lateral offset from the center line $e_{y}$, and for large inputs and large rates of change of the input, where $q_c>0$ and $\bm R,\bm R_d\geq0$. Moreover, the last term is included to maximize the progress of the EV along the track depending on $q_s>0$, where $s$ is the longitudinal position along the track, initialized based on the current state $\bm\xi_t$ of the EV~\eqref{eqn:opponent_s_init} and predicted based on the predicted longitudinal velocity $V_{x,k}$ of the EV~\eqref{eqn:opponent_s_dynamics}. The dynamics of the EV~\eqref{eqn:dynamics_opponent} is the dynamic bicycle model without compensation, as in this scenario, we focus exclusively on the uncertainty introduced by the unknown policy of the opponent, to allow a comparison with~\cite{zhu2022}. Constraints~\eqref{eqn:max_ey_opponent} and~\eqref{eqn:max_input_opponent} enforce track boundary and input constraints, respectively. Constraint~\eqref{eqn:opponent_collision_avoidance} enforces collision avoidance with the opponent, whose predicted state at step $k$ is $\bm\xi_k^\text{O}$. Collision avoidance constraints also take the uncertainty around the prediction into account, as explained in the next paragraph. To avoid embedding the GP model in the optimization problem, but still include the reaction of the opponent to the EV future movements, the GP input feature $\bm z_\text{O}$ is constructed using the open-loop solution of the EV MPC optimal control problem from the previous iteration.

\subsubsection*{\textbf{Probabilistic Collision Avoidance Constraints}}
Collision avoidance constraints~\eqref{eqn:opponent_collision_avoidance} consists of ellipsoidal regions around the predicted positions of the opponent, that the EV must not enter. At first, the minimum covering ellipse given the physical dimensions of the opponent is considered; then, the ellipse is expanded by considering the uncertainty around the prediction of the opponent in longitudinal and lateral directions. Observe that the posterior covariance provided by the GP is fundamental to expanding the forbidden ellipsoidal regions. Finally, the constraints are implemented as soft constraints~\cite[Section IV]{zhu2022}, to allow for small violations of the expanded ellipsoidal regions if this yields a significant advantage in terms of performance, although such violation is disincentivized. More details on the collision-avoidance constraints are reported in~\cite[Section IV]{zhu2022}. It is worth observing that, although the quadratic collision avoidance constraints make the optimal control problem non-convex, the solution is obtained efficiently using the FORCESPRO software~\cite{domahidi2014}.

\subsubsection*{\textbf{Active Exploration and EV Limitations}}
The exploration term of the cost function~\eqref{eqn:cost_function_opponent} is
\begin{equation}
    J_\text{expl}=\sum_{k=1}^{\Nmpc}\lVert\bm\xi_k-\bm\xi^\text{ref}_k\rVert^2_{\bm Q},
\end{equation}
that is, penalizes deviations from the (time-varying) reference state $\bm\xi^\text{ref}_k$, weighted by matrix $\bm Q\geq0$. The EV reference state is selected to test the opponent's reaction to different overtaking attempts of the EV, in order to collect meaningful data points that allow for an accurate representation of the opponent's behavior with the GP model.

The reference state $\bm\xi^\text{ref}_k$ is determined as follows. At first, we determine the reference state based on the current state of the extended system: we approximate the future relative configuration of the two vehicles using the current relative configuration. Similarly, we use the current linear velocity and yaw angle of the two vehicles as reference for the next step. Then we refine the reference state by determining the target GP input feature $\bm z^{\text{ref}}_\text{O}$ using Algorithm~\ref{alg:feature_selection}. On the one hand, the distance $d_i$ of a candidate feature from the feature $\tilde{\bm z}$ visited following performance objectives is penalized, to prevent the EV from moving in a possibly dangerous way; on the other hand, visiting regions of the feature space with high posterior covariance $v_i$ is encouraged, to test the reaction of the opponent to behaviors of the EV for which the prediction of the reaction is more uncertain. However, using the target GP input feature from Algorithm~\ref{alg:feature_selection} might not result in sufficiently diverse data. In fact, the opponent's future moves are not controlled by the EV, therefore the EV cannot arbitrarily enforce the future configuration of the two racing vehicles. There is the possibility that, while the EV is attempting to reach a given traffic configuration, the opponent reacts in a way to counterbalance the movement of the EV, and the configuration of the two reaches an equilibrium. In this case, although the absolute position of the two vehicles changes, the relative position does not change. If such equilibrium is reached and the change in the relative configuration is smaller than a threshold for several consecutive steps, we heuristically modify Algorithm~\ref{alg:feature_selection} to encourage the exploration of GP input features corresponding to configurations of the two agents that are different from the current configuration by reversing vector $\bm D_i$ in Algorithm~\ref{alg:feature_selection}, with a view to breaking the stalemate. Furthermore, the target GP input feature is not updated for a few iterations, to avoid reaching the previous equilibrium.

Given the selected target GP input feature $\bm z^{\text{ref}}_\text{O}$, the EV reference state for each prediction step $k$ is computed in order to create the traffic configuration associated with $\bm z^{\text{ref}}_\text{O}$, based on the current prediction of the opponent $\bm\xi^\text{O}_k$. It is worth observing that, other than the GP input features that depend on both the EV and the opponent, namely the longitudinal distance $s^\text{O}-s$ and the lateral distance $e_y^\text{O}-e_y$, there are several GP input features over which the EV has no control, namely the opponent yaw angle and lateral velocity and the curvature of the look-ahead points. For such features, the exploration cannot be encouraged. Therefore, when generating the list of candidate GP input features $\{\bm z^{(i)}_\text{O}\}_i^{\ngrid}$ considered in Algorithm~\ref{alg:feature_selection} it is worth considering features that differ only in the components over which the EV has influence. Otherwise, the algorithm might choose a target GP input feature $\bm z^{\text{ref}}_\text{O}$ because of the high posterior covariance given by the curvature value, for example, which the EV cannot impose, and possibly resulting in a relative configuration for which the posterior covariance is already small. Focusing only on the input features over which the EV has influence is beneficial also to maintain the number of candidate target GP input features $\ngrid$ limited.

\subsubsection*{\textbf{Iterative Framework in Head-to-Head Racing}}
Finally, we outline how our iterative scheme works for the scenario with the opponent. At first, an initial GP model is used during the exploration phase. At this stage, it is not required that the GP model is accurate since it will improve in the training after the exploration phase. Nevertheless, a GP model is needed, as the regions of the feature space that must be explored are chosen using the posterior covariance. Therefore, at first, a coarse model is trained, possibly from data of other opponents, and therefore not tailored for the current opponent.

Then, the exploration phase takes place, selecting a high value of $\alpha$ in the EV optimal control problem~\eqref{prb:opponent_problem}. The goal of this phase is a trade-off between winning the race and collecting a variety of informative data points about the opponent's policy in reaction to several attempts of the EV. At the end of the exploration phase, which can last for a few minutes, the GP is retrained on a remote platform, while the EV continues the competition. As soon as the training of the updated GP model has been completed, the updated GP model is transferred to the EV, which now can leverage an accurate prediction of the opponent's behavior and focus on winning the race, that is, setting parameter $\alpha=0$ in the optimal control problem~\eqref{prb:opponent_problem}. In theory, several iterations of the exploration phase can be run, to further increase the accuracy of the GP predictor by collecting more data for the regions of the feature space that are still not well represented. However, our simulations show that a single phase of exploration already yields a significant improvement, as we discuss in Section~\ref{sec:opponent_simulations}. Furthermore, several runs of exploration phase and retraining are only possible for sufficiently long competitions.

\section{Simulations}
\label{sec:simulations}
In this section, we describe the simulations that were conducted to validate our iterative GP regression framework with active exploration mechanisms in both autonomous racing scenarios. In the time trial, Section~\ref{sec:laptime_simulations}, we compare performance with previous work~\cite{su2023}, where a double iterative GP regression framework is used without active exploration. We show that our approach yields an improvement in the minimum lap time as a result of the more accurate learning performance that the enriched dataset from the exploration allows. Then, in the challenge with the opponent, Section~\ref{sec:opponent_simulations}, we compare our approach with the approach from~\cite{zhu2022}, in which a GP predictor of the future trajectory of the opponent is trained on a large dataset of shorter runs. Notably, our approach results in an improvement in the average EV performance as a consequence of the improved prediction of the opponent for further prediction steps, although our approach relies on a significantly smaller dataset of measurements collected during a single phase of exploration.

\subsection{Time Trial}
\label{sec:laptime_simulations}
We used the same simulation setup as in previous work~\cite{su2023} for an accurate and fair comparison. The closed-loop simulations were carried out in the highly realistic racing simulation platform Gran Turismo Sport from Sony Interactive Entertainment Inc~\cite{SonyGTS}, using as EV the Audi TT Cup running on the Tokyo Expressway Central Outer Loop Track. The desktop computer wired connected to the Play Station 5 is an Alienware-R13, with CPU Intel i9-12900 and GPU Nvidia 3090. Our code is developed in Python. The QP MPC optimal control problem~\eqref{prb:mpc_problem} is solved using \textit{CVXPY}~\cite{diamond2016} and \textit{QP solver}~\cite{domahidi2013}. The MPC frequency is $20\ \si{Hz}$ and the prediction horizon $\Nmpc=20$. The nonlinear optimal control problem for path planning~\eqref{prb:planning_problem} is solved using \textit{CasADi}~\cite{andersson2019,wachter2006}. The vehicle parameters are reported in Table~\ref{tab:vehicle_params_GTS}. The GP regression is implemented using \textit{GPyTorch}~\cite{gardner2018}, which exploits the GPU and adopts an efficient and general approximation of GPs based on black-box matrix-matrix multiplication. Over 10,000 data points are supported in the GP dataset while preserving almost the same prediction accuracy and making the GP regression estimation feasible in real-time.

\renewcommand\arraystretch{1.2} 
\begin{table}
    \centering
    \caption{Parameters of Audi TT Cup in GTS}
    \begin{tabular}{lc}
    \hline \hline
    Parameter & Value \\
    \hline
    Total mass $m$ &  1161.25 $\si{kg}$  \\
    Length from CoG to front wheel $l_\text{f}$ & 1.0234 $\si{m}$ \\
    Length from CoG to rear wheel $l_\text{r}$ & 1.4826 $\si{m}$ \\
    Width of chassis &1.983 $\si{m}$ \\
    Height of CoG $h_c$ &  0.5136 $\si{m}$ \\
    Friction ratio $\mu$ & 1.5\\
    Wind drag coefficient $C_{xw}$ & 0.1412 $\si{kg/m}$ \\
    Moment of inertia $I_{zz}$ & 2106.9543 $\si{Nm}$ \\
    \hline \hline
    \end{tabular}
    \label{tab:vehicle_params_GTS}
\end{table}
\renewcommand{\arraystretch}{1.0}

In our implementation, the maximum size of the dataset is $\ndict=2000$. The risk parameter in the tightened constraints~\eqref{eqn:ey_constraint_tightening} is $\beta=0.6$.
It is important to observe that since the target GP input feature is changed at each iteration of the MPC algorithm, we collect data points also about sudden changes in the features, which are relevant for the GP model in the planning problem~\eqref{prb:planning_problem}. To speed up the target GP input feature selection online in Algorithm~\ref{alg:feature_selection}, we evaluate and store the covariance of each point in the list of candidate features $\{\bm z^{(i)}\}_i^{\ngrid}$ before the start of each trial, after retraining the GP. During the first two iterations, we encourage the exploration setting $\alpha_0=\frac{6}{7}=0.857$ and $\alpha_1=\frac{5}{7}=0.714$. We observed that after the first two iterations of the iterative GP regression algorithm, the selected dataset of measurements is not updated further, because the collected measurements are sufficiently diverse. Therefore, from the third iteration, $\alpha$ is set to zero, instead of gradually decreasing it to zero, that is, from iteration number 2, the focus is exclusively on minimizing the lap time.

In the first run, the EV uses the nominal MPC, without the GP compensation, to track a curvature-optimal path~\cite{hao2022}, and the measurements are used to train the two GPs. In the following iterations, the GPs are exploited and the optimal path is re-planned. The planning problem is warm-started with the planned trajectory from the previous iteration, to discourage large deviations from the previous trajectory, since this could result in planning infeasible trajectories. Furthermore, we have heuristically observed that the optimal planned trajectory does not improve significantly after the first iteration of the optimization, therefore we only run one iteration. The iterative framework has been repeated for 7 iterations. Because of small uncertainties in the timing of the communication network between the computer implementing our algorithm and the simulation environment in the Play Station, slight variations are observed when the same simulation is repeated. Thus, we have repeated each simulation three times, considering the mean of the measured times and the standard deviation between the three trials.

\begin{figure*}
    \centering
    \begin{subfigure}[t]{0.3\textwidth}
         \includegraphics[width=\textwidth]{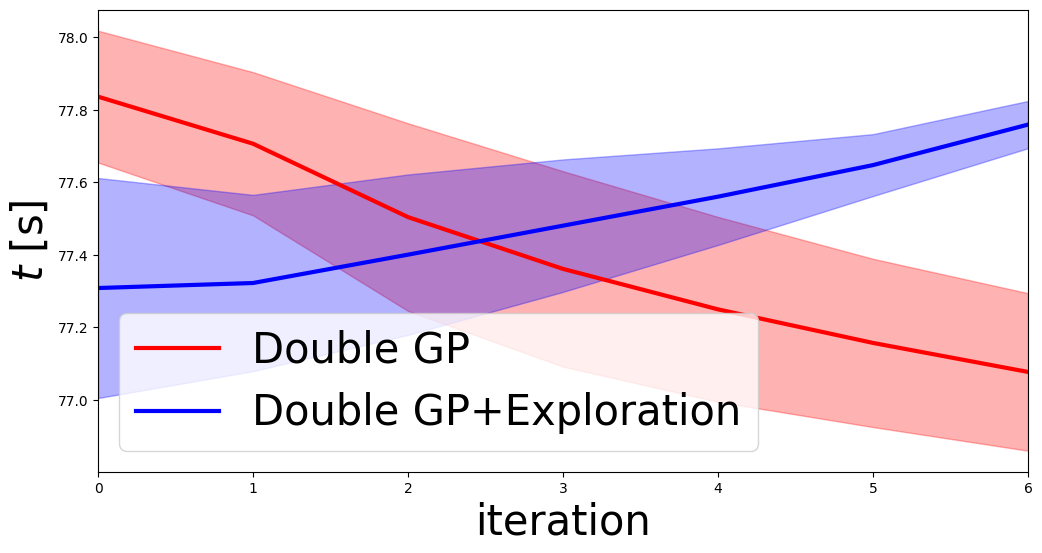}
         \caption{Time of the planned path.}
         \label{fig:lap_time_planned}
     \end{subfigure}
    \begin{subfigure}[t]{0.3\textwidth}
         \includegraphics[width=\textwidth]{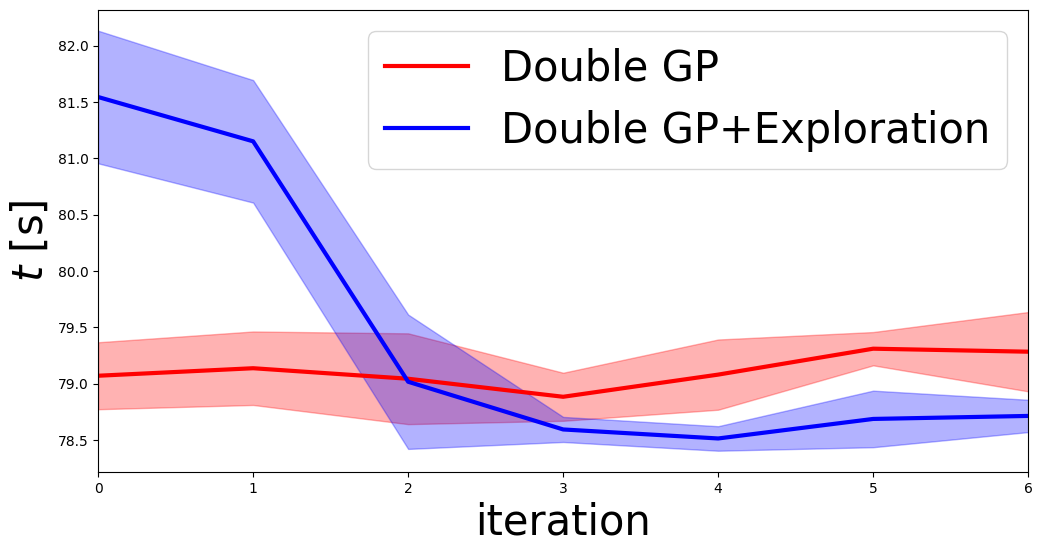}
         \caption{Time measured during the trial in each iteration.}
         \label{fig:lap_time_run}
     \end{subfigure}
    \begin{subfigure}[t]{0.3\textwidth}
         \includegraphics[width=\textwidth]{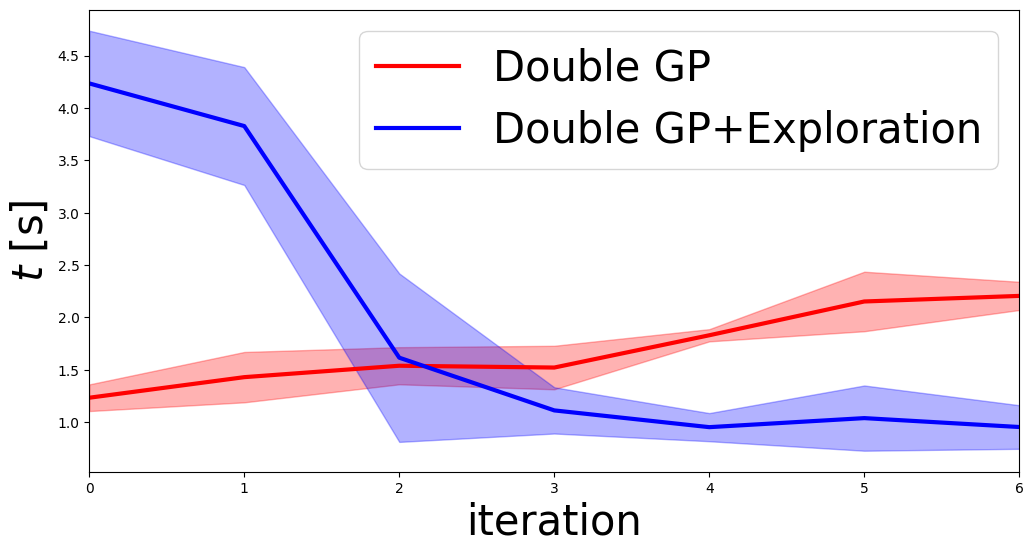}
         \caption{Difference between planned and measured.}
         \label{fig:lap_time_difference}
     \end{subfigure}
    \caption{Lap times obtained in each iteration in \textit{Time Trial} simulations in Gran Turismo Sport. Each simulation has been repeated for three trials: solid lines indicate the mean over the three trials, and the shaded areas represent the standard deviation across the three trials. At each iteration, \textbf{Double GP+Exploration }uses $\alpha_0=6/7,$ $ \alpha_1=5/7,$ and $ \alpha_{2+}=0$ to transition from high exploration to no exploration.}
    \label{fig:lap_times}
\end{figure*}

Figure~\ref{fig:lap_times} shows the lap times obtained with the comparison with the \textbf{Double GP} method, the previous approach~\cite{su2023}, and our proposed \textbf{Double GP+Exploration} which includes our active exploration method in Algorithm 1. For convenience, iteration 0 is the first trial in which the GP compensation is used, that is, neglecting the curvature-optimal run. As shown in Figure~\ref{fig:lap_time_planned}, in our approach, the time of the optimal planned path increases over the iterations, although the dispersion within repeated simulations decreases. This is understood as a consequence of the improvement in learning performance. In fact, the goal of the path planner is to derive the optimal path that is feasible for the actual dynamics of the EV, therefore it is reasonable that a more accurate dynamics results in an optimal planned path with higher lap time. Considering the actual lap time measured at each iteration, reported in Figure~\ref{fig:lap_time_run}, we first observe a significant increase in the lap time yielded by our algorithm, which is due to the fact that in the first two iterations, the exploration takes place, and therefore the performance objectives are partially compromised to collect diverse measurements. However, from iteration 2, the focus of the EV is on minimizing the lap time and we observe a decrease in the minimum time as well as in the deviation between repeated simulations compared to the baseline~\cite{su2023}. Finally, the difference between the lap time of the planned path and the actual lap time of the run, Figure~\ref{fig:lap_time_difference}, shows that the diverse dataset resulting from the active exploration reduces the gap between the time of the planned path and the actually achieved minimum time over the iterations.

\begin{figure}
    \centering
    \begin{subfigure}[c]{0.45\textwidth}
         \includegraphics[width=\textwidth]{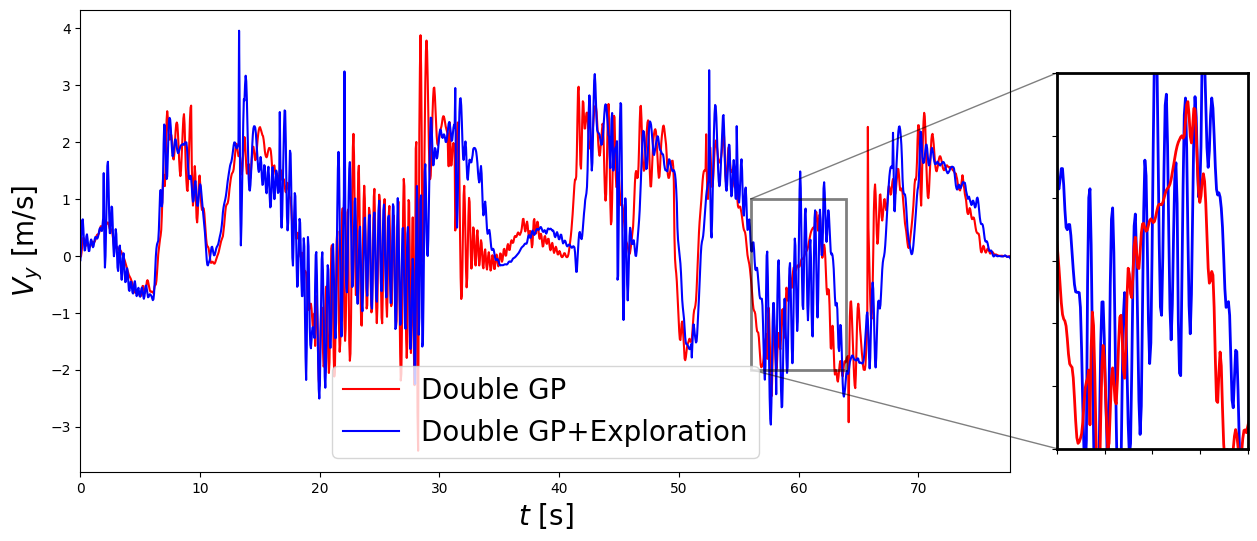}
         \caption{Data collected for $V_y$.}
         \label{fig:data_collected_Vy}
     \end{subfigure}
    \begin{subfigure}[c]{0.45\textwidth}
         \includegraphics[width=\textwidth]{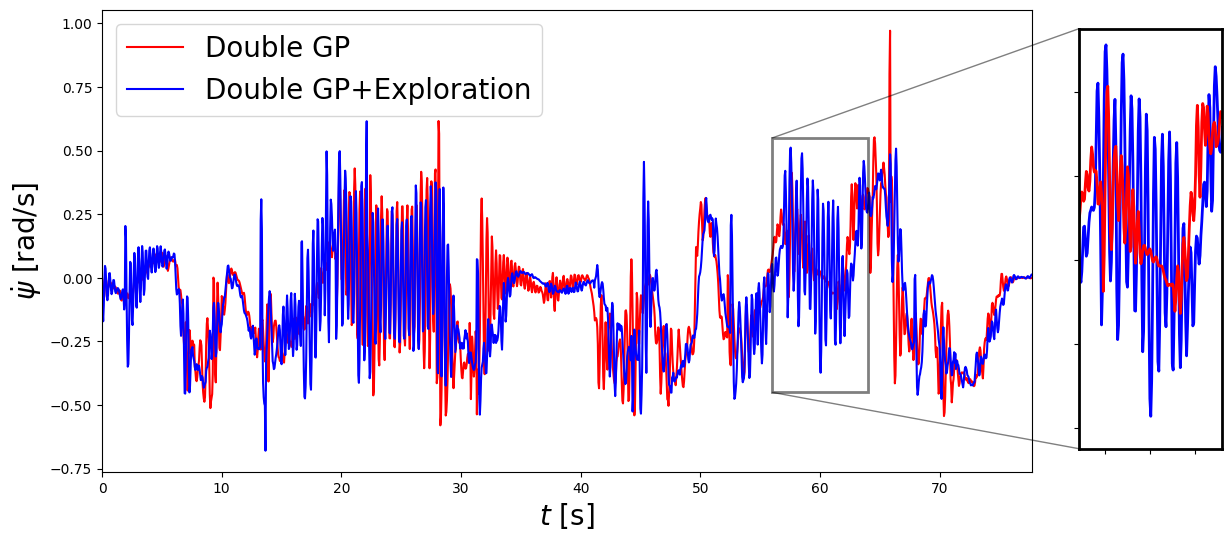}
         \caption{Data collected for $\dot{\psi}$.}
         \label{fig:data_collected_dpsi}
     \end{subfigure}

    \caption{State analysis of \textit{Time Trial} simulations in Gran Turismo Sport. Data was collected in the first iteration, in which the EV dynamics is tested by repeatedly deviating from the trajectory of the planned path ($\alpha_0=6/7$).}
     \label{fig:data_collected_first_iteration}
\end{figure}

Figure~\ref{fig:data_collected_first_iteration} presents the data collected in the first iteration. Both from the signal of the longitudinal velocity $V_y$, Figure~\ref{fig:data_collected_Vy}, and the signal of the derivative of the yaw angle $\dot{\psi}$, Figure~\ref{fig:data_collected_dpsi}, we observe that during the run the EV dynamics is tested by repeatedly deviating from the trajectory of the planned path. 

\begin{figure}
    \centering
    \begin{subfigure}[c]{0.24\textwidth}
         \includegraphics[width=\textwidth]{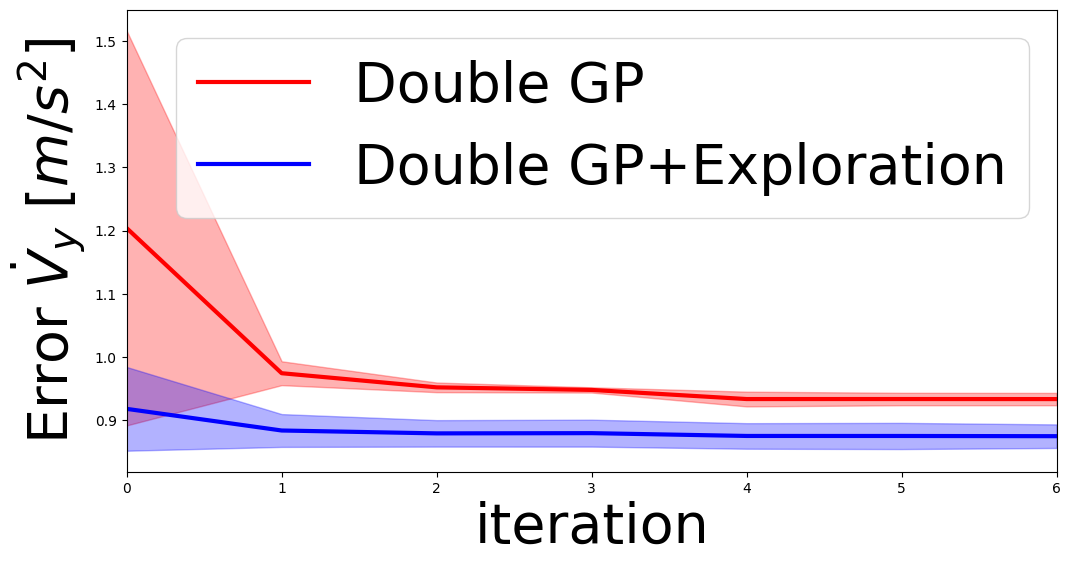}
         \caption{GP for planning, feature $\dot{V}_y$.}
     \end{subfigure}
    \begin{subfigure}[c]{0.24\textwidth}
         \includegraphics[width=\textwidth]{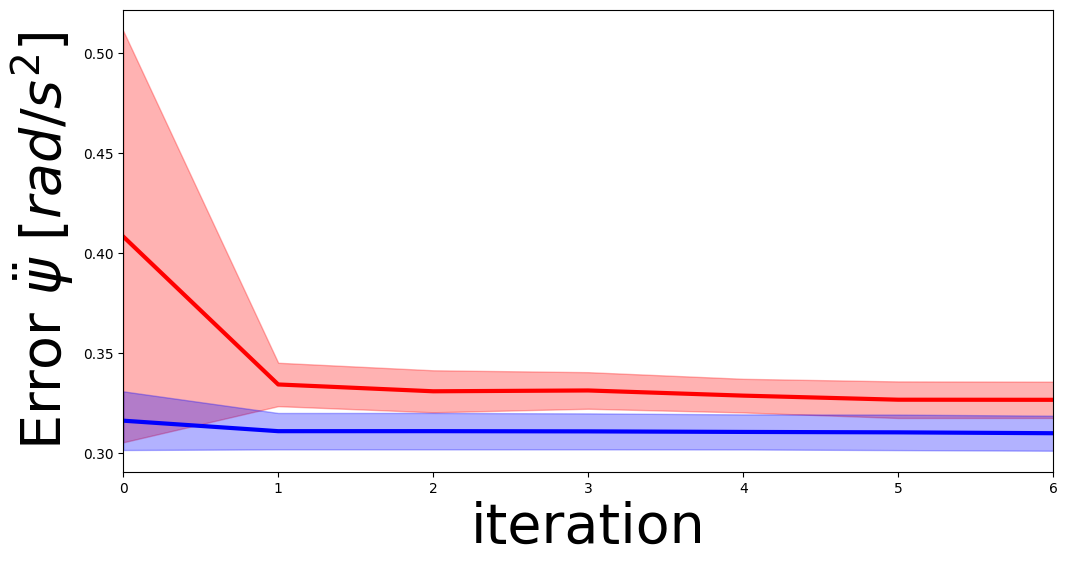}
         \caption{GP for planning, feature $\ddot{\psi}$.}
     \end{subfigure}
    \begin{subfigure}[c]{0.24\textwidth}
         \includegraphics[width=\textwidth]{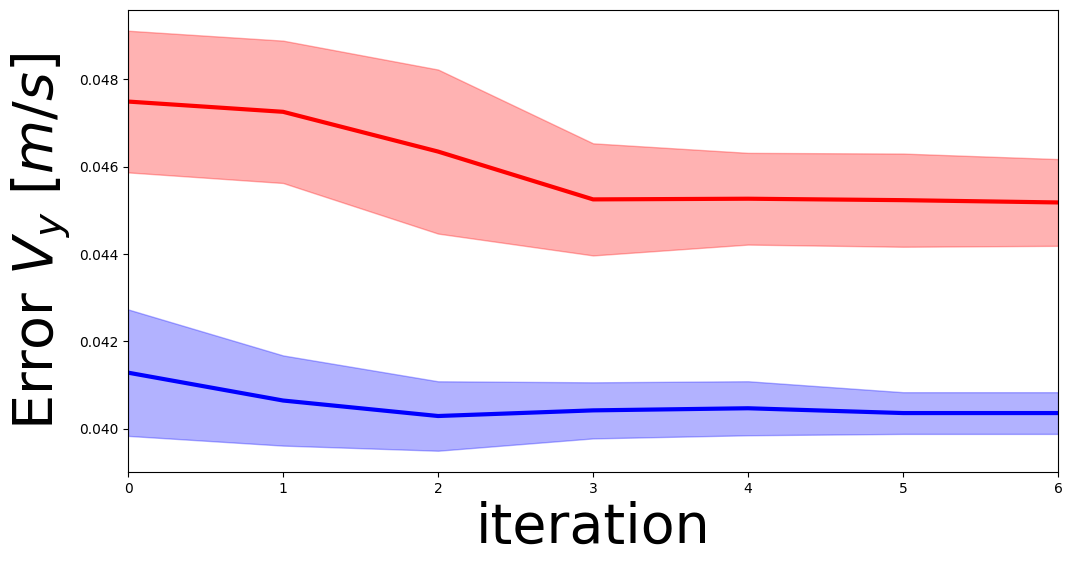}
         \caption{GP for tracking, feature $V_y$.}
     \end{subfigure}
    \begin{subfigure}[c]{0.24\textwidth}
         \includegraphics[width=\textwidth]{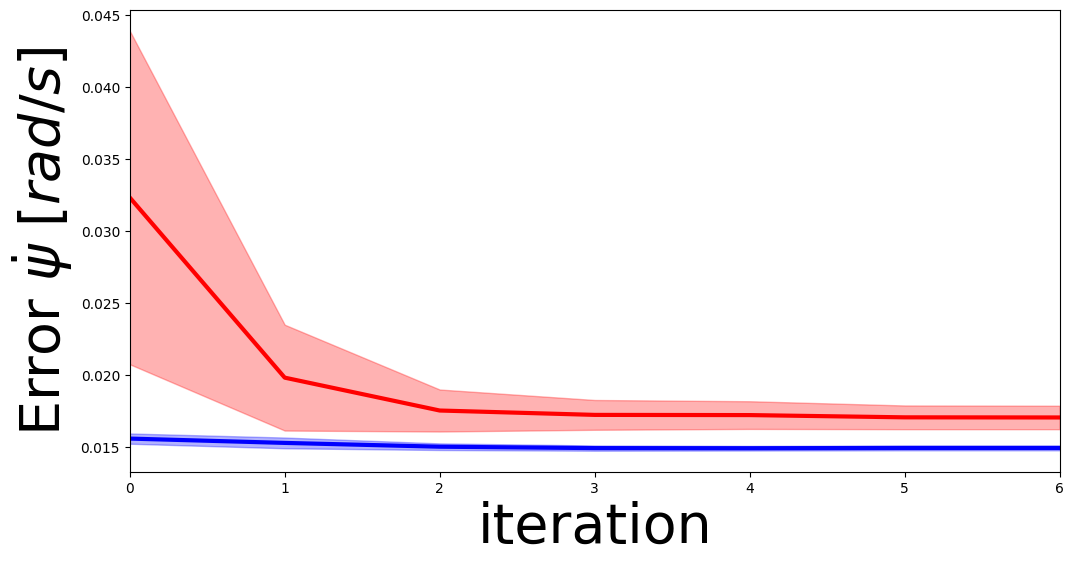}
         \caption{GP for tracking, feature $\dot{\psi}$.}
     \end{subfigure}
    \caption{Prediction error between compensated model and data of the dynamics, for each iteration in \textit{Time-Trial} task. Each simulation has been repeated for three trials: solid lines indicate the mean, and the shaded areas represent the standard deviation across the three trials.}
    \label{fig:learning_evaluation_gts}
\end{figure}

Finally, we have evaluated the learning performance over the iterations, both for the GP used in the planning and for the GP used in the tracking phase. The evaluation, shown in Figure~\ref{fig:learning_evaluation_gts}, investigates the prediction error over a dataset of diverse measurements, collected during a run in which the EV repeatedly deviated from the center line of the track. With respect to the previous work~\cite{su2023}, the diverse dataset collected during the active exploration allows a reduction of the prediction error for both the GP used in the planning and the GP used in the tracking phase, confirming that the improved performance in the measured lap time is a consequence of increased GP prediction accuracy.

\subsection{Head-to-head Racing}
\label{sec:opponent_simulations}

For the competition against the opponent, we use the simulation setup from~\cite{zhu2022}, which implements a racing environment for miniature racing cars\footnote{\href{https://github.com/MPC-Berkeley/gp-opponent-prediction-models}{https://github.com/MPC-Berkeley/gp-opponent-prediction-models}.}. The optimal control problem~\eqref{prb:opponent_problem} is formulated in \textit{CasADi}~\cite{andersson2019} and solved using sequential quadratic approximations with the QP solver \textit{hpipm}~\cite{frison2020} in the \textit{FORCESPRO} software~\cite{domahidi2014}. The control frequency is $10\ \si{Hz}$, and the prediction horizon is $\Nmpc=10$. All simulations are run on a laptop with an AMD Ryzen 5 3500U eight-core processor.

In all simulations, the opponent is implemented as an MPC-controlled agent with a blocking policy~\cite{brudigam2021d}. Other than performance-based objectives, the cost function of the opponent penalizes deviations from the current lateral position on the track of the EV, so that the opponent \lq\lq mirrors\rq\rq~the EV lateral behavior and blocks overtaking attempts. To encourage overtaking attempts of the EV, the parameter $q_s$ ruling the progress maximization reward of the EV in~\eqref{eqn:cost_function_opponent} is set higher than for the TV. Further details are given in~\cite{zhu2022}.

We compare two methods for modeling the opponent with the GP. In \textbf{Baseline}~\cite{zhu2022}, the GP model for the opponent is trained using closed-loop trajectories from an offline dataset of 500 runs in which the EV starts behind the opponent on randomly generated tracks. 
In contrast, our proposed \textbf{Data Selection + Exploration} method uses the iterative and exploration-based approach presented in Section~\ref{sec:method_opponent}. For the initial GP model, we use a \textit{smaller} initial offline dataset of 20 runs, generated with the same mechanism as the Baseline method. This allows us to test how well the GP model can be improved during the exploration. The exploration phase lasts 10 minutes of simulation time and is run on the closed track provided by~\cite{zhu2022}. During exploration, we set $\alpha=0.9$, so that the EV focuses primarily on testing the opponent's reaction to several EV movements. We simulate only one exploration and retrain phase, thus the GP model is retrained only at the end of the exploration. Then, the parameter $\alpha$ is set to zero for all following simulations, that is, the EV focuses exclusively on winning the race.

To test each method, we randomly generate a set of 100 scenarios. A track is randomly generated from straight, curved, and chicane stretches, and the length and curvature of each stretch are randomly selected. For each track, the initial longitudinal position and velocity of the cars are randomly generated, but the EV is always behind, to test the overtaking ability. Each simulation is interrupted 1.5 seconds after overtaking occurs, or when the EV reaches the track end. 

We summarize the results in Table \ref{table:head2head}. In 100 simulations, no major collision is observed. Nevertheless, the EV hits the track border in 2 simulations when the GP predictor from~\cite{zhu2022} is used, and 5 times when using our GP trained with the exploration data. For the average overtaking time, we consider the simulations on the 93 tracks in which the EV stays strictly inside the track boundaries with both predictors. On average, the EV overtakes the opponent 0.33 seconds earlier when using the GP trained on the exploration data compared to the baseline approach. It should be observed that this improvement in the EV performance is achieved with a significantly smaller training dataset, that is, roughly 600 data points collected during the exploration, as opposed to the dataset of sample runs used to train the baseline GP~\cite{zhu2022}, consisting of roughly 5000 data points. The \href{https://tumde-my.sharepoint.com/:v:/g/personal/tommaso_benciolini_tum_de/EdM3c6IwOqxIt3-Ot0pGRQABM4E0NLX82nnWa9XxGloO-A?nav=eyJyZWZlcnJhbEluZm8iOnsicmVmZXJyYWxBcHAiOiJPbmVEcml2ZUZvckJ1c2luZXNzIiwicmVmZXJyYWxBcHBQbGF0Zm9ybSI6IldlYiIsInJlZmVycmFsTW9kZSI6InZpZXciLCJyZWZlcnJhbFZpZXciOiJNeUZpbGVzTGlua0RpcmVjdCJ9fQ&e=o8FpnR}{video} shows the exploration phase in the challenge with the opponent and the transition to focusing on winning the race after the training of the GP with the enriched dataset.

\renewcommand\arraystretch{1.5}
\begin{table*}[t]
\caption{\textit{Head-to-Head} racing results over 100 tracks}
    \centering
\begin{threeparttable}
\centering
    \begin{tabular}{ccccccc}
    \hline \hline
        & & Average overtaking & \multicolumn{4}{c}{Prediction error mean$\pm$std [\si{m}] over all simulations}\\
        GP Predictor type & Hit track border & time mean$\pm$std [\si{s}] & 1-step-ahead & 2-step-ahead & 8-step-ahead & 9-step-ahead\\
    \hline
    
    \textbf{Baseline~\cite{zhu2022}} & 2 & 12.772$\pm$4.353 & {0.003$\pm$0.003} & {0.009$\pm$0.009} & {0.100$\pm$0.100} & {0.119$\pm$0.119}\\
    
    \multirow{3}{2.5cm}{\centering \textbf{Data Selection \\+\\ Exploration}}&  &  &  &  &  &  \\
    & 5 & 12.442$\pm$5.041 & {0.004$\pm$0.004} & {0.007$\pm$0.007} & {0.046$\pm$0.046} & {0.055$\pm$0.055}\\
    &  &  &  &  &  &  \\

    \textbf{Data Selection}\tnote{1} & / & / & {0.005$\pm$0.005} & {0.011$\pm$0.011} & {0.108$\pm$0.105} & {0.131$\pm$0.127}\\
    \hline \hline
\end{tabular}
\begin{tablenotes}
    \item[1] The \textit{Data Selection} GP predictor is only used for the offline analysis of the prediction accuracy. Therefore, we only report the average prediction errors computed in the offline analysis with respect to the closed-loop opponent trajectories.
\end{tablenotes}    
\end{threeparttable}
\label{table:head2head}
\end{table*}
\renewcommand{\arraystretch}{1.0}

Finally, we analyze the GP prediction accuracy to test how well the data selection and exploration mechanisms reduce model error. We compare the methods to a third GP, the \textbf{Data Selection} method, which trains a GP on a small dataset of the most diverse measurements within the dataset used for the Baseline GP, therefore without exploration. The dataset is selected using the procedure outlined in Section~\ref{sec:method}. We evaluate the impact of employing only the most diverse data points collected within several runs, without employing the active exploration mechanism in the data collection. We assess model accuracy by comparing the prediction error of the lateral position of the opponent, which is of primary concern for overtaking maneuvers. We perform the analysis offline using the 200 closed-loop trajectories of the opponent collected from the simulations on the 100 tracks in which the EV first uses the Baseline GP and then our GP with Data Selection and Exploration. Since the trajectory of the opponent depends on the EV's own behavior, we repeat the prediction offline using data from all 200 trajectories for all three GP predictors, for a fair comparison.

\begin{figure}
    \centering
    \begin{subfigure}[c]{0.38\textwidth}
         \includegraphics[width=\textwidth]{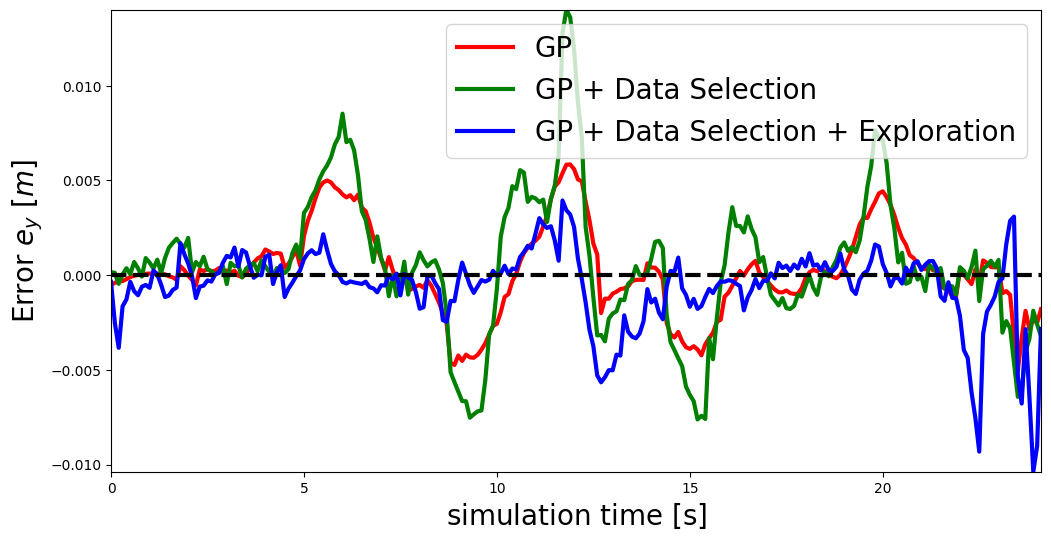}
         \caption{1-step-ahead prediction during the run.}
         \label{fig:opponent_pred_error_1_step}
     \end{subfigure}
    \begin{subfigure}[c]{0.38\textwidth}
         \includegraphics[width=\textwidth]{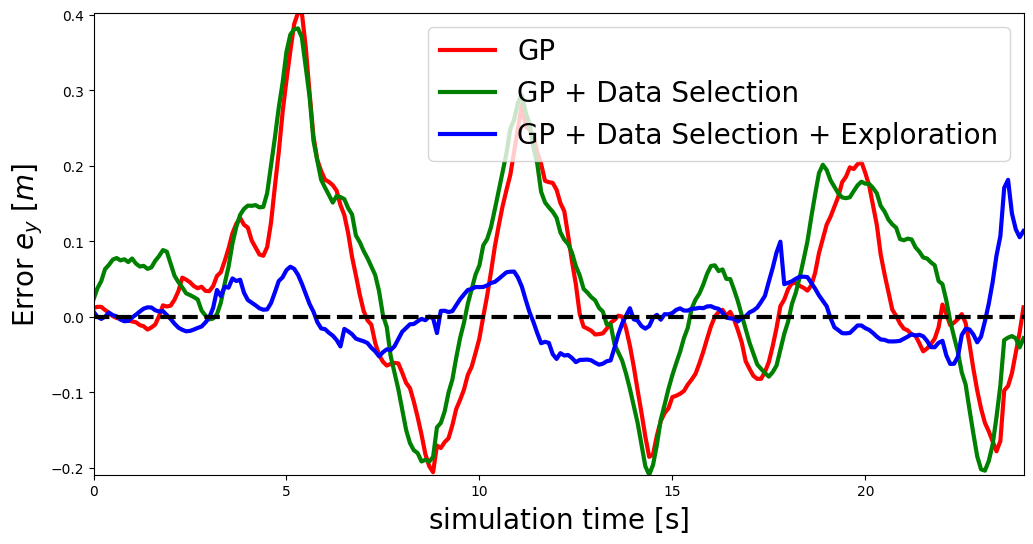}
         \caption{9-step-ahead prediction error during the run.}
         \label{fig:opponent_pred_error_9_step}
     \end{subfigure}
    \begin{subfigure}[c]{0.38\textwidth}
         \includegraphics[width=\textwidth]{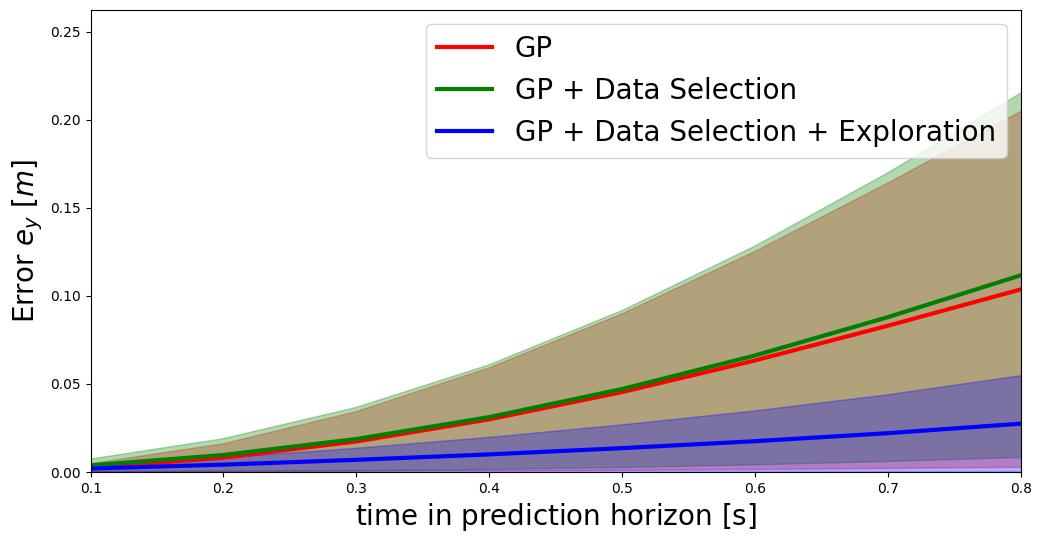}
    \caption{Average prediction error $t$-seconds-ahead. Solid lines indicate the mean, the shaded areas the standard deviation.}
     \label{fig:opponent_pred_error_for_time}
     \end{subfigure}
    \caption{Prediction error of the opponent lateral position $e_y$ with respect to the same closed-loop trajectory from one of the closed-loop trajectories opponent.
    }
    \label{fig:opponent_pred_errors}
\end{figure}

The analysis of the accuracy is shown in Figure~\ref{fig:opponent_pred_errors}, using the data from one of the 200 closed-loop opponent trajectories. Furthermore, on the right of Table \ref{table:head2head} we report the average results over all 200 closed-loop trajectories. Each GP method has a comparable accuracy in the 1-step-ahead prediction of the lateral position of the opponent, shown in Figure~\ref{fig:opponent_pred_error_1_step}. However, there are occasional spikes in prediction error, especially with the GP  only using Data Selection to improve dataset diversity. This likely indicates that exploration is necessary to improve the diversity of the dataset and, thus, the accuracy of the GP. 

While 1-step prediction accuracy is important, accurately predicting the opponent's behavior over \textit{long} time horizons is also very important, given the fact that the EV's behavior is predicted using an N-step prediction horizon in the MPC. Thus, we compare each method's 9-step prediction accuracy in Figure~\ref{fig:opponent_pred_error_9_step}. Our proposed GP with data selection and exploration significantly outperforms the other methods, resulting in a smaller 9-step prediction error compared to the two other predictors. In fact, comparing the average $t$-step prediction error in Figure~\ref{fig:opponent_pred_error_for_time} as a function of the number of steps in the horizon $t$, we see that the exploration mechanism 
decreases modeling error compared to the Baseline GP or Data Selection alone.  As expected, each method's accuracy deteriorates for further prediction steps; however, our exploration-based GP results in the slowest increase in the mean and standard deviation of prediction error as the prediction horizon increases. Since the Baseline GP and GP with Data Selection achieve similar modeling performance, this indicates that the exploration mechanism can make a notable improvement in the training dataset by purposefully opting to collect data in regions with greater modeling uncertainty. Thus, with the greater long-step prediction accuracy of the opponent's model, our GP with Data Selection and Exploration improves the performance of the EV, since the strategic decisions especially rely on the prediction for further steps.

\section{Conclusion}
\label{sec:conclusion}
In this work, we presented an iterative Gaussian Process regression scheme for autonomous racing implementing an active exploration mechanism. During the first iterations, the EV trajectory is planned trying to collect measurements for the states with high posterior covariance. Among the collected measurements, a smaller dataset is obtained, selecting the most diverse data points, and is used to retrain the Gaussian Process model. Then, in further iterations of the algorithm, the focus is exclusively on improving the performance of the EV, leveraging on the improved prediction accuracy. We showed that the GP exploration method can be applied both when the GP model is used for error compensation and for opponent modeling. We tested the framework to compensate for the modeling errors in the EV dynamics near handling limits in Gran Turismo Sport~\cite{SonyGTS}, and to model the opponent's reaction to the EV's own decisions in a simulation environment. In both scenarios, we obtained a significant improvement in the prediction accuracy and, consequently, in the EV performance.

Future research will focus on the validation of the active exploration approach for the opponent challenge in other simulation environments, with a special focus on the generalizability of the approach to different opponent policies. Furthermore, the framework will be tested in scenarios in which both uncertainties, namely modeling errors in the vehicle dynamics and the opponent policy, are tackled simultaneously.

Furthermore, it is worth discussing strategies to determine when the exploration phase must be interrupted in the opponent challenge. Evaluating the improved accuracy of the GP prediction is not possible in real-time, because new training is required. One possibility is to interrupt the exploration depending on how diverse the dataset of collected measurements is, for example, terminating the exploration when the dataset reaches a steady state because the newly collected measurements are sufficiently well covered by past ones.

\section*{Acknowledgments}
The authors thank Shaoshu Su for valuable discussions, and Ce Hao for the simulation setup in GTS. We also would like to thank Kenta Kawamoto from Sony AI for his kind help and fruitful discussions. This work was supported by Sony AI, and Polyphony Digital Inc., which provided the Gran Turismo Sport framework. T. Benciolini's visit to the University of California, Berkeley, was supported by a fellowship within the \lq\lq Research Grants for Doctoral Students\rq\rq~program of the German Academic Exchange Service (DAAD) and the Bavaria California Technology Center (BaCaTeC) grant 12-[2022-2]. Catherine Weaver is supported by the National Science Foundation Graduate Research Fellowship Program under Grant No. DGE 1752814. Any opinions, findings, conclusions, or recommendations expressed in this material are those of the authors and do not necessarily reflect the views of the National Science Foundation.

\appendix
\section*{Planning Optimal Control Problem}

We transform the integral in time into an integral in the spatial domain~\cite{kapania2016} and closely approximate the solution that would be obtained by directly considering the lap time as objective, as in~\cite{hao2022}, where a more complete discussion is provided. The time-optimal trajectory is obtained by solving
\begin{subequations}
\begin{equation}
    \min_{\{\bm u_k\}_{k=0}^{\Np-1}} \sum_{k=0}^{\Np-1}\frac{(1-\kappa(s_k)e_{y,k})\Delta s_k}{V_{x,k}\cos(e_{\psi,k})-V_{y,k}\sin(e_{\psi),k}}\label{eqn:cost_function_planning}
\end{equation}
\begin{alignat}{2}
    \text{s.t. }\bm \xi_{k+1}&=\bm \xi_k+\bm{f}(\bm \xi_k,\bm u_k)T+&&\bm\mu^\text{plan}(\bm z_\text{plan}(\hat{\bm\xi}_k, \hat{\bm u}_k)),\nonumber\\
    &\ &&\forall k=0,\dots,\Np-1\label{eqn:dynamics_path_planning}\\
    \bm\xi_{\Np} &= \bm\xi_0\label{eqn:closed_path_planning}\\
    w_{\text{r},k}&\leq e_{y,k}\leq w_{\text{l},k},\ &&\forall k=1,\dots,\Np\label{eqn:max_ey_planning}\\
    \delta_{\text{min},k}&\leq \delta_k\leq \delta_{\text{max},k},\ &&\forall k=0,\dots,\Np-1,\label{eqn:max_delta_planning}
\end{alignat}
\label{prb:planning_problem}%
\end{subequations}
where $\Np$ is the length of the horizon for the planning problem. Cost function~\eqref{eqn:cost_function_planning} results from the transformation of the time-optimal objective into an integral in the spatial domain, as in~\cite{hao2022}, and $\Delta s_k$ is the incremental longitudinal progress along the path. Constraint~\eqref{eqn:closed_path_planning} enforces that the path starts and ends in the same point, whereas~\eqref{eqn:max_ey_planning} and~\eqref{eqn:max_delta_planning} guarantee that the trajectory does not leave the right and left track boundaries $w_{\text{r}}$ and $w_{\text{l}}$ and that the steering angle $\delta$ remains in the actuation range $[\delta_{\text{min}},\delta_{\text{max}}]$. Constraint~\eqref{eqn:dynamics_path_planning} guarantees that the planned trajectory is feasible for the vehicle dynamics, which is crucially important to ensure that the vehicle can track the optimal trajectory. $\bm\mu^\text{plan}(\bm z_\text{plan}(\hat{\bm\xi}_k, \hat{\bm u}_k))$ represents the modeling error compensation provided by the GP model, correcting the inaccuracies of the physics-based model when the vehicle is driven at handling limits. We model the dynamics compensation as a GP
\begin{equation}
    \bm g^\text{plan}(\bm z_\text{plan})\sim\mathcal{N}\left(\bm\mu^\text{plan}(\bm z_\text{plan}),\bm\Sigma^\text{plan}(\bm z_\text{plan})\right).
\end{equation}
We compensate the two states with the greatest impact on the prediction error, namely $\dot{V_y}$ and $\ddot{\psi}$~\cite{su2023}, therefore $\bm\mu^\text{plan}(\bm z_\text{plan})=[0,\mu^\text{plan}_{\dot{V_y}}(\bm z_\text{plan}),\mu^\text{plan}_{\ddot{\psi}}(\bm z_\text{plan}),0,0,0]^\top$~\cite{hewing2018}. As in~\cite{su2023}, we consider as input features of the planning GP $\bm g^\text{plan}$ vector $\bm z_\text{plan}=[V_y,\dot{\psi},\delta]^\top$, being the most correlated with the output features. In order not to embed the GP model into the optimization problem, the input feature $\bm z_\text{plan}$ is defined with respect to nominal state and input vector $\tilde{\bm\xi}_k$ and $\tilde{\bm u}_k$, that is, the solution of the previous iteration of the optimization~\cite{hao2022}, rather than on the actual state and input.

\bibliographystyle{IEEEtran}
\bibliography{res}

\end{document}